\documentclass[12pt]{article}

\usepackage{graphicx}
\usepackage{subcaption}
\usepackage{epsfig}
\usepackage{epstopdf}

\usepackage[T1]{fontenc}

\DeclareSymbolFont{myletters}{OML}{ztmcm}{m}{it}
\DeclareMathSymbol{\uplambda}{\mathord}{myletters}{"15}
\DeclareMathSymbol{\upxi}{\mathord}{myletters}{"18}

\usepackage{amsfonts}

\usepackage{lmodern} 
\usepackage{amsmath}                                                    
\usepackage{amsthm}                                                     
\usepackage{amssymb}
\usepackage{multirow}
\numberwithin{equation}{section} 
\newcommand{\newc}{\newcommand}
\newc{\be}{\begin{equation}}
\newc{\ee}{\end{equation}}
\newc{\bg}{\begin{gathered}}
\newc{\eg}{\end{gathered}}
\newc{\tref}[1]{Table \ref{#1}}
\newc{\eref}[1]{Equation \eqref{#1}}
\newc{\su}[1]{$SU(#1)$}
\newc{\bm}[1]{\mathbf{#1}}
\newc{\fref}[1]{Figure \ref{#1}}

\newc{\ra}{\rightarrow}
\newc{\lra}{\leftrightarrow}
\newc{\ov}{\overline}
\newc{\ba}{\begin{eqnarray}}
\newc{\ea}{\end{eqnarray}}
\newc{\mf}{\mathsf}

\DeclareMathOperator{\tr}{tr}

\begin{document}
\begin{titlepage}

\vspace*{0.7cm}

\begin{center}
{
\bf\large
  Inflation from a no-scale supersymmetric $SU(4)_{C}\times{SU(2)_{L}\times{SU(2)_{R}}}$ model}
\\[12mm]
Waqas Ahmed$^{a }$ \footnote{E-mail: \texttt{waqasmit@itp.ac.cn}},   Athanasios Karozas$^{b }$
\footnote{E-mail: \texttt{akarozas@cc.uoi.gr}}

\end{center}

\vspace*{0.50cm}
	\centerline{$^{a}$ \it
		CAS Key Laboratory of Theoretical Physics, }
	\centerline{\it
		Institute of Theoretical Physics, Chinese Academy of Sciences, }
		\centerline{\it
		Beijing 100190, P. R. China}
	\vspace*{0.2cm}
	\centerline{$^{b}$ \it
		Physics Department, Theory Division, Ioannina University,}
	\centerline{\it
		GR-45110 Ioannina, Greece}
		\vspace*{0.2cm}
	\vspace*{1.20cm}
\vspace*{1.20cm}

\begin{abstract}
\noindent
We study  inflation in a supersymmetric Pati-Salam model  driven by a potential generated in the context of no-scale supergravity. The Pati-Salam gauge group  $SU(4)_{C}\times{SU(2)_{L}\times{SU(2)_{R}}}$, is supplemented  with a $Z_{2}$ symmetry. Spontaneous breaking via the $SU(4)$ adjoint leads to the left-right symmetric group. Then the $SU(2)_{R}$ breaks at an intermediate scale and the inflaton is a combination of the neutral components of the $SU(2)_R$ doublets. We discuss various limits of the parameter space and we show that consistent solutions with the cosmological data for the  spectral index $n_{s}$ and the tensor-to-scalar ratio $r$ are found for a wide range of the parameter space of the model. Regarding the latter, which is a canonical measure of primordial gravity waves, we find that $r\sim{10^{-3}-10^{-2}}$.  An alternative possibility where the adjoint scalar field $S$ has the r\^ole of the inflaton is also discussed.

  \end{abstract}

  \end{titlepage}

\thispagestyle{empty}
\vfill
\newpage

\setcounter{footnote}{0}

\section{INTRODUCTION }

In  cosmological models inflation is realized  by a slowly rolling scalar field, the so called  inflaton, whose energy density  dominates  the early history Universe \cite{Guth:1980zm,Linde:1981mu,Mukhanov:1981xt,Albrecht:1982wi}. 
Among  several suggestions regarding its origin, the economical scenario  that  this field can be identified with the 
Standard Model (SM) Higgs state $\mathrm{h}$, has received considerable attention\cite{Bezrukov:2007ep}.  In this approach, the
 Higgs field  drives inflation  through its strong coupling,  $\upxi \mathrm{h}^2 R$,  where $R$ is the Ricci scalar and $\upxi$
 is a dimensionless parameter that acquires a large value,  $\upxi\gtrsim 10^4$.

In  modern particle physics theories,  cosmological  inflation is usually described within  the framework of supergravity or 
superstring grand unified theories (GUTs). In these theories the SM is embedded in a higher gauge symmetry and the field content including 
the  Higgses  are incorporated in representations of the higher symmetry which includes the SM gauge group.  In this context, 
several new facts and constraints should be taken into account. For instance,  since new symmetry breaking stages are involved,  
the Higgs sector is usually  extented and  alternative possibilities for identifying the inflaton emerge. In addition, the effective 
potential has a specific structure  constrained from fundamental principles of the theory. In string theory effective models, for
 example, in a wide class of compactifications the scalar potential  appears with a no-scale structure  as in standard supergravity 
theories \cite{Cremmer:1983bf, Lahanas:1986uc}.   In general, the scalar potential is a function of the various fields which enter in a complicated 
manner through the superpotential $W$ and the K\"ahler potential $K$.  Thus, a  rather detailed investigation is required to determine 
the conditions for slow roll inflation and  ensure a stable  inflationary trajectory in such models. Modifications of the basic no-scale K\"ahler potential and various choices for the superpotential have been
studied leading to a number of different inflationary cases \cite{Ellis:2013xoa}-\cite{Romao:2017uwa}, while studies of inflation within supergravity in a model independent way can be found in \cite{Covi:2008cn, Hardeman:2010fh}.

In the present work we implement the scenario of Higgs inflation in a model based on the Pati-Salam gauge symmetry $SU(4)_{C}\times SU(2)_L
\times SU(2)_R$ \cite{Pati:1974yy} (denoted for brevity with 4-2-2). This model has well known  attractive features (see for example the 
recent review \cite{Pati:2017ysg}) and has been successfully rederived  in superstring and D-brane 
theories \cite{Antoniadis:1988cm, Cvetic:2004ui, Anastasopoulos:2010ca, Cvetic:2015txa}. Early universe cosmology and inflationary predictions of the model (or its extensions) have been discussed previously in several works \cite{Jeannerot:2000sv, Pallis:2011gr, Bryant:2016tzg}.  Here we consider a supersymmetric version of the 4-2-2 model where the breaking down to the SM gauge group takes place in two steps. First $SU(4)$ breaks  
spontaneously at the usual supesymmetric GUT scale $M_{GUT}\gtrsim 10^{16}$ GeV, down to the \emph{left-right} group\footnote{For a recent discussion on left-right models based on GUTs, see \cite{Chakrabortty:2017mgi}. Inflation from an $SO(10)$ model with left-right intermediate symmetry is analysed in \cite{Garg:2015mra}.} via the adjoint representation. Then,  depending on the specific structure of the Higgs
sector, the $SU(2)_R$ scale can break either at the GUT scale, i.e., simultaneously  with $SU(4)$, or at some lower, intermediate energy scale. 
The variety of possibilities are reflected back to the effective field theory model  implying  various interesting phenomenological 
consequences. Regarding the Higgs inflation scenario, in particular, the inflaton field can be identified with the neutral components of the $SU(2)_{R}$ doublet fields
associated with the intermediate scale symmetry breaking.  In this work we will explore alternative  possibilities to realise inflation 
where the inflaton is identified with the $SU(2)_{R}$ doublets. We also examine the case of inflation in the presence of  the adjoint representation.

The layout of the paper is as follows. In section 2, we present a brief description of the 4-2-2 model, focusing in its particle content
and the  symmetry breaking pattern. In sections 3 we present the  superpotential and the emergent no-scale supergavity K\"ahler potential
of the effective model. We derive the effective potential and analyse the predictions on inflation when either the $SU(2)_{R}$ doublets or the adjoint play the r\^ole of the inflaton. We present our conclusions in section 4.

\section{DESCRIPTION OF THE MODEL}
In this section we highlight the basic ingredients of  the model with gauge symmetry,
\be \label{psgroup}
 SU(4)_{C}\times{SU(2)_{L}}\times{SU(2)_{R}}~\cdot
 \ee
 \noindent This model unifies each family of quarks and leptons into two irreducible representations,
  $F_{i}$ and $\bar{F}_{i}$  transforming as \cite{King:1997ia}
 \[F_{i}=(4,2,1)_{i}\quad{\text{and}}\quad \bar{F}_{i}=(\overline{4},1,2)_{i}~,\]

\noindent under the corresponding factors of the gauge group~(\ref{psgroup}). Here the subscript $i$ ($i=1,2,3$) denotes family index. 
 Note that $F+\bar{F}$ comprise the $16$ of $SO(10)$,   $16\rightarrow{(4,2,1)+(\overline{4},1,2)}$.
The explicit embedding  of the SM matter fields, including the right-handed neutrino is as follows:	
\be
F_{i}=
\begin{pmatrix} 
u_r & u_g & u_b & \nu  \\
d_r & d_g & d_b & e
\end{pmatrix}_{i}\quad{,}\quad{\bar{F}_{i}=
\begin{pmatrix} 
u^{c}_r & u^{c}_g & u^{c}_b & \nu^{c}  \\
d^{c}_r & d^{c}_g & d^{c}_b & e^{c}
\end{pmatrix}_{i}}~,
\ee

\noindent where the subscript $(r,g,b)$ are color indices.

 The  symmetry breaking 
\be
 SU(4)_{C}\times{SU(2)_{R}}\rightarrow{SU(3)_{C}\times{U(1)_{Y}}}~,
\ee

\noindent is achieved by introducing two Higgs multiplets 

\be\label{HiggsofPS}
H=(\overline{4},1,2)=
\begin{pmatrix} 
u_{H}^{c} & u_{H}^{c} & u_{H}^{c} & \nu_{H}^c  \\
d_{H}^{c} & d_{H}^{c} & d_{H}^{c} & e_{H}^{c}
\end{pmatrix}\quad{,}\quad{\bar{H}=(4,1,2)=
\begin{pmatrix} 
\ov{u}_{H}^{c} & \ov{u}_{H}^{c} & \ov{u}_{H}^{c} & \ov{\nu}_{H}^c  \\
\ov{d}_{H}^{c} & \ov{d}_{H}^{c} & \ov{d}_{H}^{c} & \ov{e}_{H}^{c}
\end{pmatrix}}
\ee
 which descend from the $16$ and $\overline{16}$ of $SO(10)$ respectively.

An alternative way to break the gauge symmetry arises in the case where the adjoint 
scalar $\Sigma=(15,1,1)$ is included in the spectrum.
 We parametrise $\Sigma$  with a singlet scalar field  $S$ 
\ba 
\Sigma\equiv (15,1,1)  &=&
\frac{S}{2\sqrt{3}}\left(
	\begin{array}{cccc}
		1&0 & 0 &0\\
		0 &1& 0&0 \\
		0 & 0 &  1&0\\
		0 & 0 &0& -3
	\end{array}
	\right)~,\label{Adj}
\ea
which acquires a GUT scale vacuum expectation value (vev) $\langle{S}\rangle\equiv\upsilon\simeq{3\times{10^{16}}}$ GeV 
 breaking  $SU(4)\to SU(3)\times U(1)$. The breaking leads to the left-right symmetric group, 
 $SU(3)_{C}\times{SU(2)_{L}}\times{SU(2)_{R}}\times{U(1)_{B-L}}$, and the decomposition of the Higgs fields  $H, \bar{H}$ 
 is as follows:
\begin{eqnarray}\label{Hbreaking}
\begin{split}
H(\ov{4},1,2)&\rightarrow{Q_{H}(\ov{3},1,2)_{-1/3}+L_{H}(1,1,2)_{1}}\\
\bar{H}(4,1,2)&\rightarrow{\ov{Q}_{H}(3,1,2)_{1/3}+\ov{L}_{H}(1,1,2)_{-1}}
\end{split}
\end{eqnarray}

\noindent where $Q_{H}=(u_{H}^{c}\quad{d_{H}^{c}})^{T}$, $\ov{Q}_{H}=(\ov{u}_{H}^{c}\quad{\ov{d}_{H}^{c}})$  
and $L_{H}=(\nu_{H}^{c}\quad{e_{H}^{c}})^{T}$, $\ov{L}_{H}=(\ov{\nu}_{H}^{c}\quad{\ov{e}_{H}^{c}})$. 

The right-handed doublets $L_{H},\ov{L}_{H}$, acquiring vev's along their neutral components ${\nu}_{H}^c ,
 \ov{\nu}_{H}^c $ and as a result they break the
$SU(2)_R$ symmetry at some scale $M_R$. This way we obtain the symmetry breaking pattern~\cite{Anastasopoulos:2010ca}:
\[
SU(4)_{C}\times{SU(2)_{R}}\times{SU(2)_{L}}\rightarrow{SU(3)_{C}\times{U(1)_{B-L}}}\times{SU(2)_{R}}\times{SU(2)_{L}}\to 
{SU(3)}\times{SU(2)_{L}}\times{U(1)_{Y}}.
\]
 The two scales $M_{GUT}$ and $M_R$ are not related to each other and it is in principle possible to 
 take $M_R$ at some lower scale provided there is no conflict with observational data such as 
 flavour changing neutral currents and lepton or baryon number violation.  
Regarding the fast proton decay problem, in particular, in 4-2-2 models,  due to absence of the associated gauge bosons
there are no contributions from dimension six (d-6) operators, and related issues from d-5 operators can be remedied with 
appropriate symmetries in the superpotential.

The remaining spectrum and its $SO(10)$ origin is as follows:  The decomposition of the $10$ representation of $SO(10)$, 
gives  a bidoublet and a sextet field, transforming under  the 4-2-2 symmetry as follows

\be 
10\rightarrow{h(1, 2, 2)+D_{6}(6,1,1)}~\cdot \label{10toHD}
\ee 

\noindent
The two Higgs doublets of the minimal supersymmetric standard model (MSSM) descend from the bidoublet

\be
h=(1,2,2)=
\begin{pmatrix}
h_{2}^{+} & h_{1}^{0}\\
h_{2}^{0} & h_{1}^{-}
\end{pmatrix}.
\ee

\noindent Also, the sextet of (\ref{10toHD}) decomposes into a pair of coloured triplets: $D_{6}\rightarrow{D_{3}(3,1,1)+\overline{D}_{3}(\ov{3},1,1)}$.

Collectively we have the following SM assignments:

\begin{equation}
\begin{split}
F&=(4,2,1)\rightarrow Q(3,2,\frac{1}{6})+L(1,2,-\frac{1}{2})\\
\bar{F}&=(\ov{4},1,2)\rightarrow u^{c}(\ov{3},1,-\frac{2}{3})+d^{c}(\ov{3},1,\frac{1}{3})+e^{c}(1,1,1)+\nu^{c}(1,1,0)\\
h&=(1,2,2)\rightarrow H_{u}(1,2,\frac{1}{2})+H_{d}(1,2,-\frac{1}{2})\\
H&=(\ov{4},1,2)\rightarrow u^{c}_{H}(\ov{3},1,-\frac{2}{3})+d^{c}_{H}(\ov{3},1,\frac{1}{3})+e^{c}_{H}(1,1,1)+\nu^{c}_{H}(1,1,0)\\
\bar{H}&=(4,1,2)\rightarrow \ov{u}^{c}_{H}(3,1,\frac{2}{3})+\ov{d}^{c}_{H}(3,1,-\frac{1}{3})+\ov{e}^{c}_{H}(1,1,-1)+\ov{\nu}^{c}_{H}(1,1,0)\\
D_{6}&=(6,1,1)\rightarrow{D_{3}(3,1,-\frac{1}{3})+\overline{D}_{3}(\ov{3},1,\frac{1}{3})}
\end{split}
\end{equation}

Fermions receive Dirac type masses from a common tree-level invariant term, $F\bar{F}h$, whilst right-handed (RH) neutrinos receive heavy Majorana contributions from
non-renormalisable terms, to be discussed in the next sections. In addition, the colour triplets $d_{H}^{c}$ and $\ov{d}_{H}^{c}$ are combined with the $D_{3}$ and $\ov{D}_{3}$ states via the trilinear operators $HHD_{6}+\bar{H} \bar{H}D_{6}$ and  get masses near the GUT scale.

After the short description of the basic features of the model, in the following sections we  investigate various inflationary scenarios in the context of no-scale supergravity, by applying the techniques presented in \cite{Ellis:2014dxa, Ellis:2016spb}.

\section{INFLATION IN NO SCALE SUPERGRAVITY}

In this section we consider  the 4-2-2 model as an  effective string theory model  and study the implications of  Higgs inflation. 
 The `light' spectrum in these constructions contains the MSSM states in representations transforming non-trivially under the 
 gauge group and a number of moduli fields associated with the particular compactification. We will focus on the superpotential and the K\"ahler potential which are essential for the study of inflation. 

The   superpotential is a holomorphic function of the fields. Ignoring Yukawa interaction terms,  the most general superpotential up to dimension four which is relevant to our discussion is 


 \begin{eqnarray}
 \begin{split}\label{wscalar}
 W&=M\bar{H}H + \mu\bar{h}h + m \tr(\Sigma)^{2}+n \bar{H}\Sigma H+ c\tr\left(\Sigma^{3}\right)  \\
 &-\alpha \left(\bar{H} H\right)^{2}-\beta\left(\bar{h}h\right)^{2} -\beta '\left(\bar{H}H\right)\left(\bar{h}h\right)- \kappa \tr\left(\Sigma^{4}\right)-\lambda \bar{H} \tr(\Sigma^{2})H
 \end{split}
 \end{eqnarray}

\noindent where from now on  we set the reduced Planck mass to unity, $M_{Pl}=1$.  We focus on the dynamics of inflation during the first symmetry breaking stages at high energy scales.  For this reason we ignore all the terms involving the bi-doubled since this state mostly contribute in low energies by ginving mass to the MSSM particles and do not play an important r\^ole during inflation. In addition we impose a $Z_{2}$ symmetry, under which $\Sigma$ is odd and all the other fields are even. As a result the trilinear terms $\bar{H}\Sigma H$ and  $\tr\left(\Sigma^{3}\right)$ are eliminated from the superpotential in (\ref{wscalar}). The elimination of these trilinear terms of the superpotential is important, since if we use $\bar{H}\Sigma H$ and  $\tr\left(\Sigma^{3}\right)$ instead of $\bar{H} \tr(\Sigma^{2})H$ and $\tr\left(\Sigma^{4}\right)$, the shape of the  resulting potential is not appropriate and it leads to inconsistent results with respect to the cosmological bounds while at the same time returns a low scale value for the parameter $M$ in the superpotential, which usually expected to be close to the GUT scale.  Then, using (\ref{Adj}) and (\ref{Hbreaking}) the superpotential takes the following form: 
 
 \begin{eqnarray}
 \begin{split}\label{superpotential2}
 W &\supset \left(M-\frac{\tilde{\lambda}}{9}S^{2}\right)\ov{Q}_{H}Q_{H}+\left(M-\tilde{\lambda}S^{2}\right)\ov{L}_{H}L_{H}-\alpha (\ov{Q}_{H}Q_{H}+\ov{L}_{H}L_{H})^{2}+mS^{2}-\tilde{\kappa}S^{4}\\
 & \quad\qquad
 \end{split}
 \end{eqnarray}
 
 \noindent where $\tilde{\lambda}=\frac{3\lambda}{4}$ and $\tilde{\kappa}=\frac{7\kappa}{12}$. From the phenomenological point of view we expect $\langle{S}\rangle=v$ to be at the GUT scale. By assuming $v\simeq{3\times{10^{16}}}$GeV and using the minimization condition $\partial{W}/\partial{S}=0$, we  estimate  that $m\simeq{2\tilde{\kappa}v^{2}}$ which, for $\tilde{\kappa}=1/2$, gives $m\sim{10^{14}}$ GeV.
 
 In the two step breaking pattern that we consider here, $\ov{L}_{H}$ and $L_H$ must remain massless at this scale in order to break the $SU(2)_R$ symmetry at a lower scale. The $SU(2)_{R}$ breaking scale should not be much lower than the  GUT scale in order to have a realistic heavy Majorana neutrino scenario. In addition we have to ensure that the coloured triplets $\ov{Q}_{H}$ and $Q_{H}$ will be heavy. In order to keep the $\ov{L}_{H}$, $L_H$ doublets at a lower scale, and at the same time the coloured fields $\ov{Q}_{H}$ and $Q_{H}$ to be heavy, we assume that $M\thickapprox{\tilde{\lambda}\langle{S}\rangle^{2}}=\tilde{\lambda}\upsilon^{2}$. In this case $\ov{Q}_{H}$, $Q_{H}$ acquire GUT scale masses  $M_{Q_{H}}\thickapprox{\frac{8\tilde{\lambda}}{9}\langle{S}\rangle^2}$.

 During inflation the colored triplets $\ov{Q}_{H}$, $Q_{H}$ and the charged components of the RH doublets, $\ov{L}_{H}$ and $L_H$, do not play an important r\^ole. The $SU(2)_R$ symmetry breaks via the neutral components\footnote{Here and for the rest of the paper, for shorthand we remove the subscript "c" on the fields, i.e: $\ov{\nu}^{c}_{H}$, $\nu^{c}_{H}\rightarrow{\ov{\nu}_{H}, \nu_{H}}$.} $\ov{\nu}_{H}$ and $\nu_{H}$. In terms of these states the superpotential reads:
 
 \begin{eqnarray}
 \begin{split}
 W= \tilde{\lambda}\left(\upsilon^{2} -S^{2}\right)\ov{\nu}_{H}\nu_{H}-\alpha (\ov{\nu}_{H}\nu_{H})^{2}+mS^{2}-\tilde{\kappa}S^{4}
 \end{split}
 \end{eqnarray}
 
\noindent where we have made use of the relation $M\simeq\tilde{\lambda}\upsilon^{2}$.

\noindent  The K\"{a}hler potential  has a no-scale structure and is a hermitian function of the  fields and their conjugates. For the present 
  analysis, we will consider the dependence of the Higgs fields of the 4-2-2 gauge group and  the `volume' modulus $T$. 
Therefore, assuming the fields $\phi_i=(S, T, H, h)$ and their complex conjugates,  we write
\begin{equation}\label{kahler1}
\begin{split}
K = -3 \log \left[T + T^{\ast}- \frac{1}{3}\left(H H^{\ast} + \bar{H} \bar{H}^{\ast} +\tr\Sigma^{\dagger}\Sigma\right) +\frac{\xi}{3}\left(H \bar{H} + H^{\ast} \bar{H}^{\ast}\right)+ \frac{\zeta}{3}\left(h h^{\ast}+\bar{h} \bar{h}^{\ast}\right)\right]
\end{split}
\end{equation}

\noindent where $\xi$ is a dimensionless parameter. In the expression (\ref{kahler1}), we can ignore the last term which involves the bidoublet and in terms of $\nu_{H}$, $\ov{\nu}_{H}$ and $S$, the K\"{a}hler potential reads: 
\begin{equation}\label{kahler2}
\begin{split}
K = -3 \log \left[T + T^{\ast}- \frac{1}{3}\left(|\nu_{H}|^{2} + |\ov{\nu}_{H}|^{2} +S^{2}\right) +\frac{\xi}{3}\left(\ov{\nu}_{H}\nu_{H} +(\ov{\nu}_{H})^{\ast}(\nu_{H})^{\ast} \right)\right].
\end{split}
\end{equation}
In order to  determine  the effective potential we define the function
\[ G= K+\log|W|^2\equiv  K+\log W+\log W^*.
\]
Then the effective potential  is given by 
\ba 
V=e^G\left(G_iG_{i j^*}^{-1}G_{j^*}-3\right)+V_D\label{VGK}
\ea 
where  $G_i (G_{j^*})$ is the derivative with respect to the field $\phi_i
(\phi^*_j)$ 
and the indices $i,j$  run over the various fields. $V_D$ stands for the D-term contribution. 

\noindent   Computing the derivatives and substituting	in  (\ref{VGK}) the  potential takes the form 
		
\begin{eqnarray}\label{fullpotential}
\begin{split}
V[\ov{\nu}_{H},\nu_{H},S]&=\frac{9}{(-3+\nu_{H}^{2}+\ov{\nu}_{H}^{2}+S^{2}-2\xi\ov{\nu}_{H}\nu_{H})^{2}}\left[(\tilde{\lambda}\upsilon^{2}-2\alpha \nu_{H}\ov{\nu}_{H})^{2}(\nu_{H}^{2}+\ov{\nu}_{H}^{2})-8\tilde{\lambda}mS^{2}\ov{\nu}_{H}\nu_{H}\right.\\
&-2\tilde{\lambda} S^{2}(\tilde{\lambda}\upsilon^{2}-2\alpha \nu_{H}\ov{\nu}_{H})(\nu_{H}^{2}+\ov{\nu}_{H}^{2})+4\tilde{\lambda}^{2}S^{2}(\ov{\nu}_{H}\nu_{H})^{2}\\
&\left. +4m^{2}S^{2}-16\tilde{\kappa}S^{4}(m-\tilde{\lambda}\ov{\nu}_{H}\nu_{H})+\tilde{\lambda}^{2}S^{4}(\nu_{H}^{2}+\ov{\nu}_{H}^{2})+16\tilde{\kappa}^{2}S^{6}\right]
\end{split}
\end{eqnarray}

\noindent where we have ignored the D-term contribution and we have assumed that the value of the $T$ modulus field is stabilized at $\langle{T}\rangle=\langle{T^{*}}\rangle=1/2$, see  \cite{Cicoli:2013rwa, Ellis:2013nxa}. Notice that in the absence of the Higgs contributions in the K\"ahler 
potential, the effective potential is exactly zero, $ V=0$ due to the well known property of the no-scale structure.

 We are going now to investigate two different inflationary cases: firstly, along H-direction and secondly along S-direction.

\subsection{INFLATION ALONG $H$-DIRECTION}

 We proceed by parametrizing the neutral components of the $L_{H}$ and $\ov{L}_{H}$ fields as $\nu_H=\dfrac{1}{2}\left(X+Y\right)e^{i\theta}$ and $\ov{\nu}_H=\dfrac{1}{2}\left(X-Y \right) e^{i\varphi}$,  respectively.  These  yield

\begin{equation}
X  = \mid \nu_H \mid + \mid \bar\nu_{H} \mid,
 \qquad Y  =  \mid \nu_H \mid - \mid \bar\nu_{H} \mid~\cdot
\end{equation}

\begin{figure}[t!]
 	\begin{subfigure}{.5\textwidth}
 		\centering
 		\includegraphics[width=.95\linewidth]{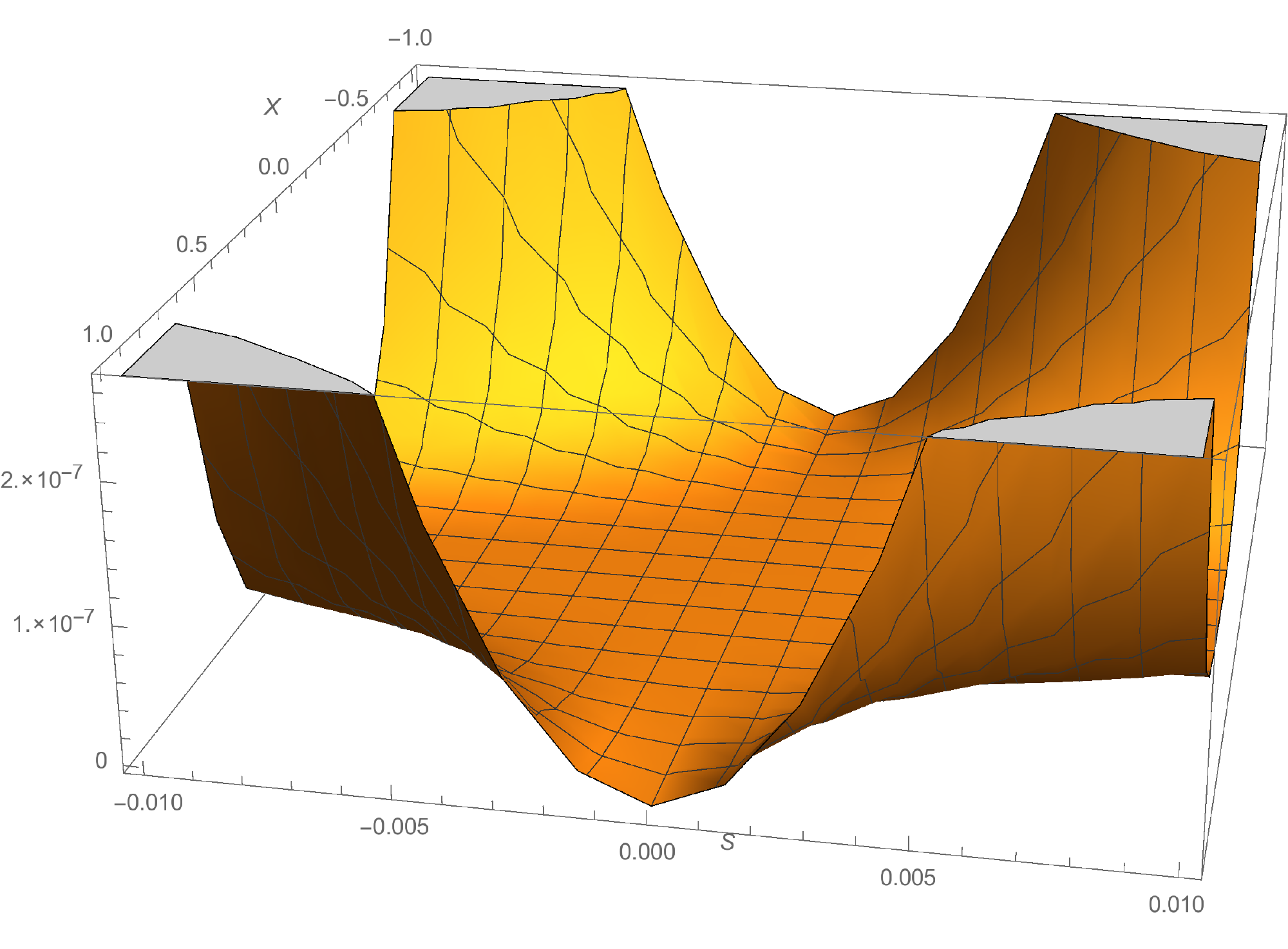}
 		\label{3d1}
 	\end{subfigure}%
 	\begin{subfigure}{.5\textwidth}
 		\centering
 		\includegraphics[width=.95\linewidth]{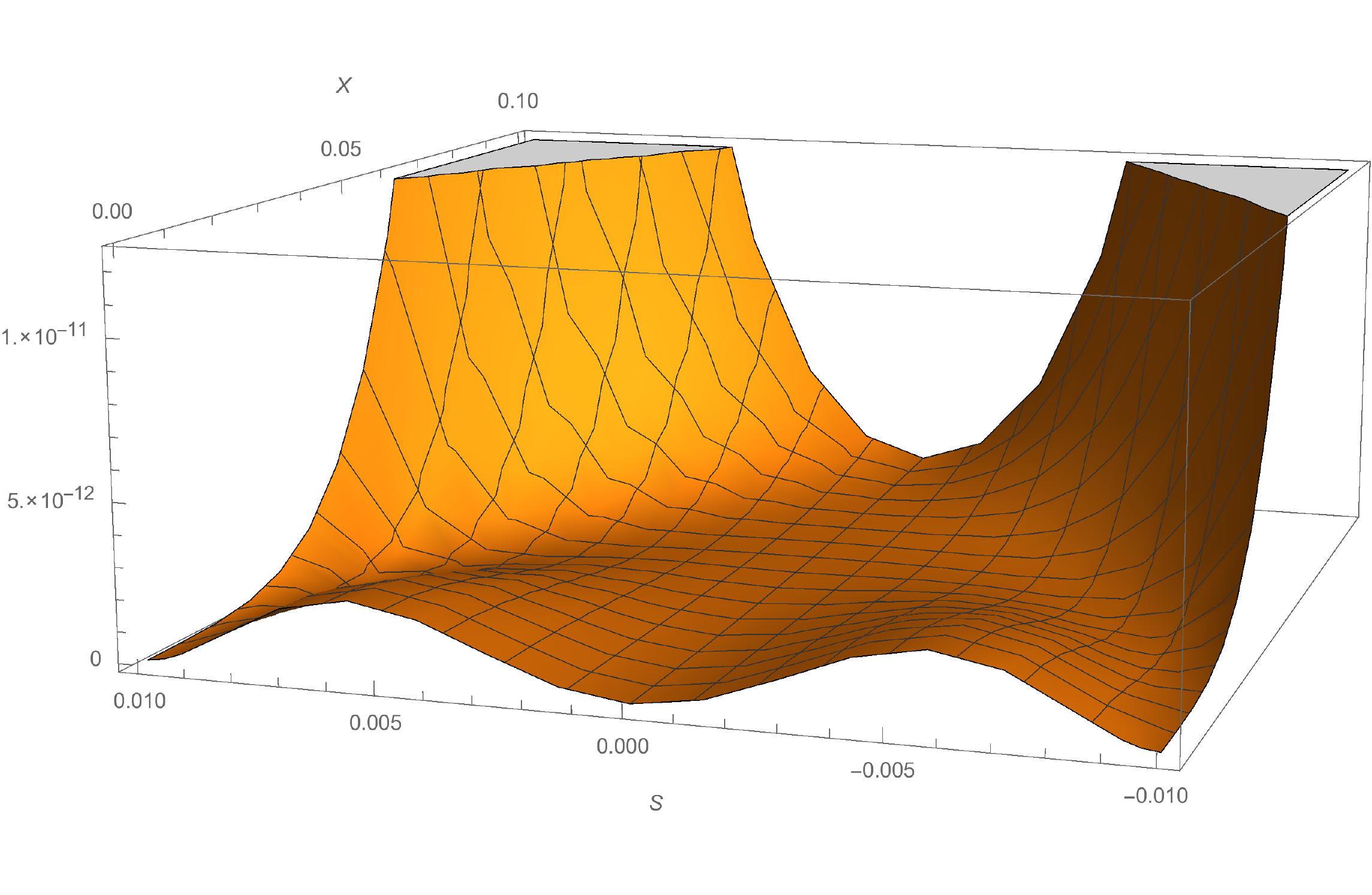}
 		\label{3d2}
 	\end{subfigure}	
 	\caption{\small{Plots of the potential as a function of $S$ and $X$ and for appropriate values of the other parameters. The plot on the right displays a close-up view  of the  region with small values for $X$ and $S$. }}
 	\label{3Dplots}
 \end{figure}

\noindent 
Assuming  $\theta=0$ and $\varphi=0$, along the D-flat direction,  $Y=0$, and the combination $X$ is identified with the inflaton. The shape of the potential, as a function of the fields $S$ and $X$, is presented in Figure \ref{3Dplots}. In order to avoid singularities from the denominator we have assume a condition which is described in the following.

 The potential along the $S=0$ direction is:
\begin{equation}\label{potential_3_6}
V\left(X\right) =\frac{\tilde{\lambda}^{2}\upsilon^{4}X^{2}\left(1-\frac{\alpha X^{2}}{2 \tilde{\lambda}\upsilon^{2}}\right)^{2}}{2\left(1-\left(\frac{1-\xi}{6}\right)X^{2}\right)^{2}} .
\end{equation}

The shape of the $V(X,S)$ scalar potential presented in Figure \ref{3Dplots} along with the inflaton trajectory description and the simplified form in (\ref{potential_3_6}) is similar with the one presented in \cite{Ellis:2014dxa,Ellis:2016spb}. As it is usually the case in no-scale supergravity,  the effective potential displays a singularity when the denominator vanishes. The presence of these singularities lead to an exponentially steep potential which can cause violation of the basic slow-roll conditions (i.e. $\varepsilon\ll{1}$, $|\eta|\ll{1}$). Consequently, these singularities must be removed. In our specific model described by the potential \eqref{potential_3_6},
 we first notice that for the special value $\xi=1$  the potential is free from singularities. For generic values of $\xi$ however, i.e. $\xi\ne 1$, the potential displays a singularity for $X=\sqrt{\frac{6}{1-\xi}}$.  In order to  remove the zeros of the denominator in \eqref{potential_3_6}, we assume the following condition \cite{Ellis:2014dxa},
\ba
\alpha=\frac{\left(1-\xi\right)\tilde{\lambda}\upsilon^{2}}{3}~\cdot\label{singularitycondition}
\ea

\noindent This is a strong assumption which relates parameters with different origins. Indeed, $\alpha$ is a superpotential parameter while $\xi$ descents from the Kahler potential. Since in our specific model the condition \eqref{singularitycondition} lacks an explanation from first principles, it will be reasonable in the subsequent analysis to study the effects of a slightly relaxed version of \eqref{singularitycondition}. This can be achieved by introducing a small parameter $\delta$ (with $\delta\ll{1}$) and modifying the condition  as follows,

\ba
\label{singularitycondition2}
\alpha=\frac{\left(1-\xi+\delta\right)\tilde{\lambda}\upsilon^{2}}{3}~\cdot \label{dsingularitycondition}
\ea

\noindent In the remaining of this section, we are going to study the potential for special $\xi$ values using the conditions~(\ref{singularitycondition}) and (\ref{dsingularitycondition}). 


We will start by analysing some special cases first. By imposing (\ref{singularitycondition}), which means $\delta=0$ 
the scalar potential simplifies to a quadratic monomial,

\begin{equation}\label{quadraticform}
V\left(X\right) = \frac{\tilde{\lambda}^{2}\upsilon^{4}}{2}X^{2}
\end{equation}
\noindent something that can be also seen from the  plots  in Figure \ref{3Dplots}, where for small values of S (along the $S=0$ direction) the potential receives a quadratic shape form. The equation (\ref{quadraticform}) shows the potential of a chaotic inflation scenario. However, at this stage,
 the inflaton field $X$ is not canonically normalized since its kinetic energy terms take the following form
\begin{equation}
\begin{split}
\mathcal{L}\left(X\right)= \frac{ 1-\frac{\xi}{6}\left(1-\xi\right) X^{2}}{2\left(1-\frac{1}{6}\left(1-\xi\right) X^{2}\right)^{2}} \left(\partial X \right)^{2} -\frac{\tilde{\lambda}^{2}\upsilon^{4}}{2}X^{2} .
\end{split}
\end{equation}
We introduce a  canonically normalized field $\chi$ satisfying 
\begin{equation}
\begin{split}
\left(\frac{d\chi}{dX}\right)^{2} = \frac{ 1-\frac{\xi}{6}\left(1-\xi\right) X^{2}}{\left(1-\frac{1}{6}\left(1-\xi\right) X^{2}\right)^{2}}.
\end{split}
\end{equation}
After integrating, we obtain the canonically normalized field $\chi$ as a function of $X$

\begin{equation}\label{hfield}
\chi  =\sqrt{6}\tanh^{-1}\left(\frac{\left(1 - \xi\right)X}{\sqrt{6\left(1-\frac{\xi\left(1-\xi\right)X^{2}}{6} \right)}}\right)
-\sqrt{\frac{6 \xi}{1-\xi}}\sin^{-1}\left(\sqrt{\xi \left(\frac{1-\xi}{6}\right)}X\right).
\end{equation}

 \noindent Next, we  investigate the implications of equation (\ref{hfield}) by considering two different cases, for $\xi=0$ and $\xi\neq{0}$. 

$\bullet$ For $\xi=0$ we have $X=\sqrt{6}\tanh\left(\frac{\chi}{\sqrt{6}}\right)$ and the potential becomes,

\begin{equation}\label{Tpotential}
V= 3 \tilde{\lambda}^{2}\upsilon^{4}  \tanh^{2}\left(\frac{\chi}{\sqrt{6}}\right),
\end{equation}
\noindent which is analogous to the conformal chaotic inflation model (or T-Model) \cite{Kallosh:2013xya}.
 In these particular type of models the potential has the general form:

\be\label{Tmodels}
 V(\chi)=\uplambda^{n}\tanh^{2n}\left(\frac{\chi}{\sqrt{6}}\right) \quad\text{where}\quad n=1,2,3,...
 \ee
 As we can see, for  $n=1$ we receive our result in (\ref{Tpotential}) with  $\uplambda=3\tilde{\lambda}^{2}\upsilon^{4}$. This potential  can be further reduced to subcases depending upon the value of $\chi$. For $\chi\geqslant1$ the potential in equation (\ref{Tpotential}) reduces to Starobinsky model \cite{Starobinsky:1980te}. In this case the inflationary observables have values $\left(n_{s},r\right)\approx \left(0.967,0.003\right)$ and the tree level prediction for $\xi=0$ is consistent with the latest {Planck} bounds \cite{Ade:2015lrj}. This type of models will be further analysed  in the next section where inflation along the $S$-direction is discussed. \\

$\bullet$ The particular case of $\xi=1$ implies a quadratic chaotic inflation and the tree-level inflationary prediction  $\left(n_{s},r\right)\approx \left(0.967,0.130\right)$ is ruled out according to the latest \emph{Planck} $2015$ results. For $0<\xi<1$ , the prediction for $\left(n_{s},r\right)$, can be
 worked out numerically.  

After this analysis we turn our attention to a numerical calculation. In our numerical analysis we imply the modified condition (\ref{singularitycondition2}) were as mentioned previously a small varying parameter $\delta$ has been introduced in order to soften the strict assumption \eqref{singularitycondition}. By substitute the relaxed condition \eqref{singularitycondition2} in \eqref{potential_3_6} and neglecting $\mathcal{O}(\delta^{2})$, the potential receives the following form:

\begin{equation}\label{potentiladelta}
V(X)\simeq{\frac{\tilde{\lambda}^{2}\upsilon^{4}}{2}X^{2}}\left(1-\frac{2\delta X^{2}}{6+(\xi-1)X^{2}}\right).
\end{equation}

\noindent As we observe the first term in the above relation is the quadratic potential \eqref{quadraticform}, while the second term encodes the effects of the small parameter $\delta$. In addition, we note that the order of the singularity enhancement have been improved in comparison with the initial potential \eqref{potential_3_6}. Next we present our numerical results where the r\^ole of the parameter $\delta$ is also discussed.

\subsection{NUMERICAL ANALYSIS }

Before presenting numerical predictions of the model it is useful to briefly review here the basic results of the slow roll assumption. The inflationary slow roll parameters are given by \cite{DeSimone:2008ei, Okada:2010jf}:
\begin{equation}
\epsilon=\dfrac{1}{2}\left(\frac{V^{\prime}\left(X\right)}{V(X)\chi^{\prime}\left(X\right)}\right)^{2} \quad{,}\quad \eta=\left(\frac{V^{\prime\prime}\left(X\right)}{V(X)\left(\chi^{\prime}\left(X\right)\right)^{2}}-\frac{V^{\prime}\left(X\right)\chi^{\prime\prime}\left(X\right)}{V(X)\left(\chi^{\prime}\left(X\right)\right)^{3}}\right).
\end{equation}
The third slow-roll parameter is,
\begin{equation}
 \varsigma^{2}=\left(\frac{V^{\prime}\left(X\right)}{V(X)\chi^{\prime}\left(X\right)}\right)\left(\frac{V^{\prime\prime\prime}\left(X\right)}{V(X)\left(\chi^{\prime}\left(X\right)\right)^{3}}-3\frac{V^{\prime\prime}\left(X\right)\chi^{\prime\prime}\left(X\right)}{V(X)\left(\chi^{\prime}\left(X\right)\right)^{4}}+3\frac{V^{\prime}\left(X\right)\left(\chi^{\prime\prime}\left(X\right)\right)^{2}}{V(X)\left(\chi^{\prime}\left(X\right)\right)^{5}}-\frac{V^{\prime}\left(X\right)\chi^{\prime\prime\prime}\left(X\right)}{V(X)\left(\chi^{\prime}\left(X\right)\right)^{4}}\right)
\end{equation}
\noindent where a prime denotes a derivative with respect to $X$. The slow-roll approximation is valid as long as the conditions $\epsilon\ll1$,$\mid \eta\mid\ll1$ and $\varsigma^{2}\ll1$
 hold true. In this scenario the tensor-to-scalar ratio $r$, the scalar spectral index $n_{s}$ and the running of the spectral index $\frac{dn_{s}}{d\ln k}$ are given by
 \begin{equation}
r\simeq16 \epsilon \quad{,}\quad n_{s}\simeq 1+2\eta-6\epsilon \quad{,}\quad \frac{dn_{s}}{d\ln k}\simeq 16\epsilon\eta-24\epsilon^{2}+2\varsigma^{2}.
 \end{equation}
 The number of e-folds is given by,
 \begin{equation}
 N_{l}=\int_{X_{e}}^{X_{l}}\left(\frac{V\left(X\right)\chi^{\prime}\left(X\right)}{V^{\prime}(X)}\right) dX,
 \end{equation}
\noindent  where $l$ is the comoving scale after crossing the horizon, $X_{l}$ is the field value at the comoving scale and $X_{e}$ is the field when inflation ends, i.e $max\left(\epsilon\left(X_{e}\right),\eta\left(X_{e}\right),\varsigma\left(X_{e}\right)\right)=1.$\\
Finally, the amplitude of the curvature perturbation $\Delta_{R}$
 is given by:
  \begin{equation}
\Delta_{R}^{2}=\frac{V\left(X\right)}{24 \pi^{2} \epsilon\left(X\right)}.
  \end{equation}

\begin{figure}[t!]
 	\begin{subfigure}{.5\textwidth}
 		\centering
 		\includegraphics[width=.9\linewidth]{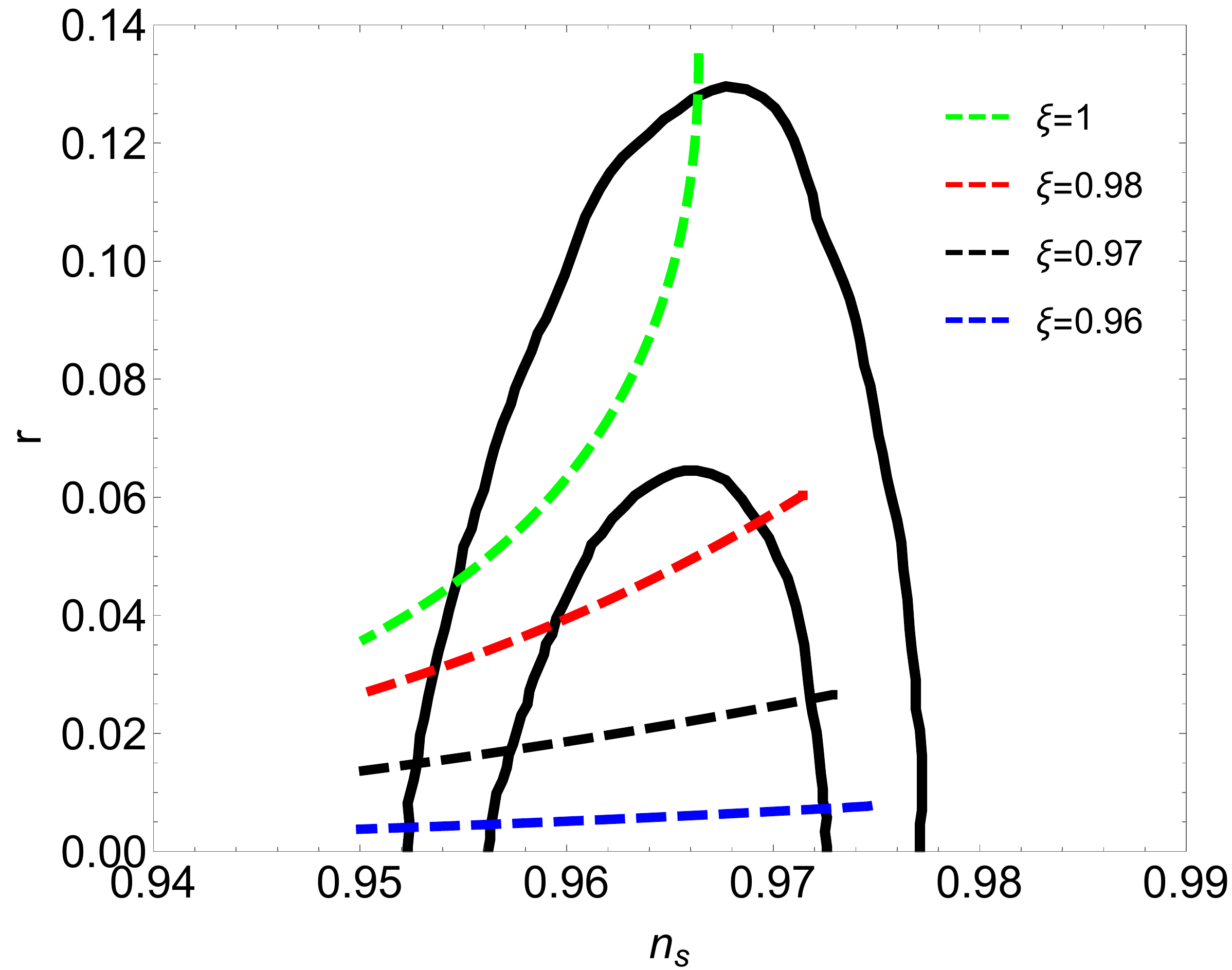}
 		\caption{\small{r vs $n_{s}$ for fixed values of $\xi$}}
 		\label{ns_r_fixed_xi}
 	\end{subfigure}%
 	\begin{subfigure}{.5\textwidth}
 		\centering
 		\includegraphics[width=.9\linewidth]{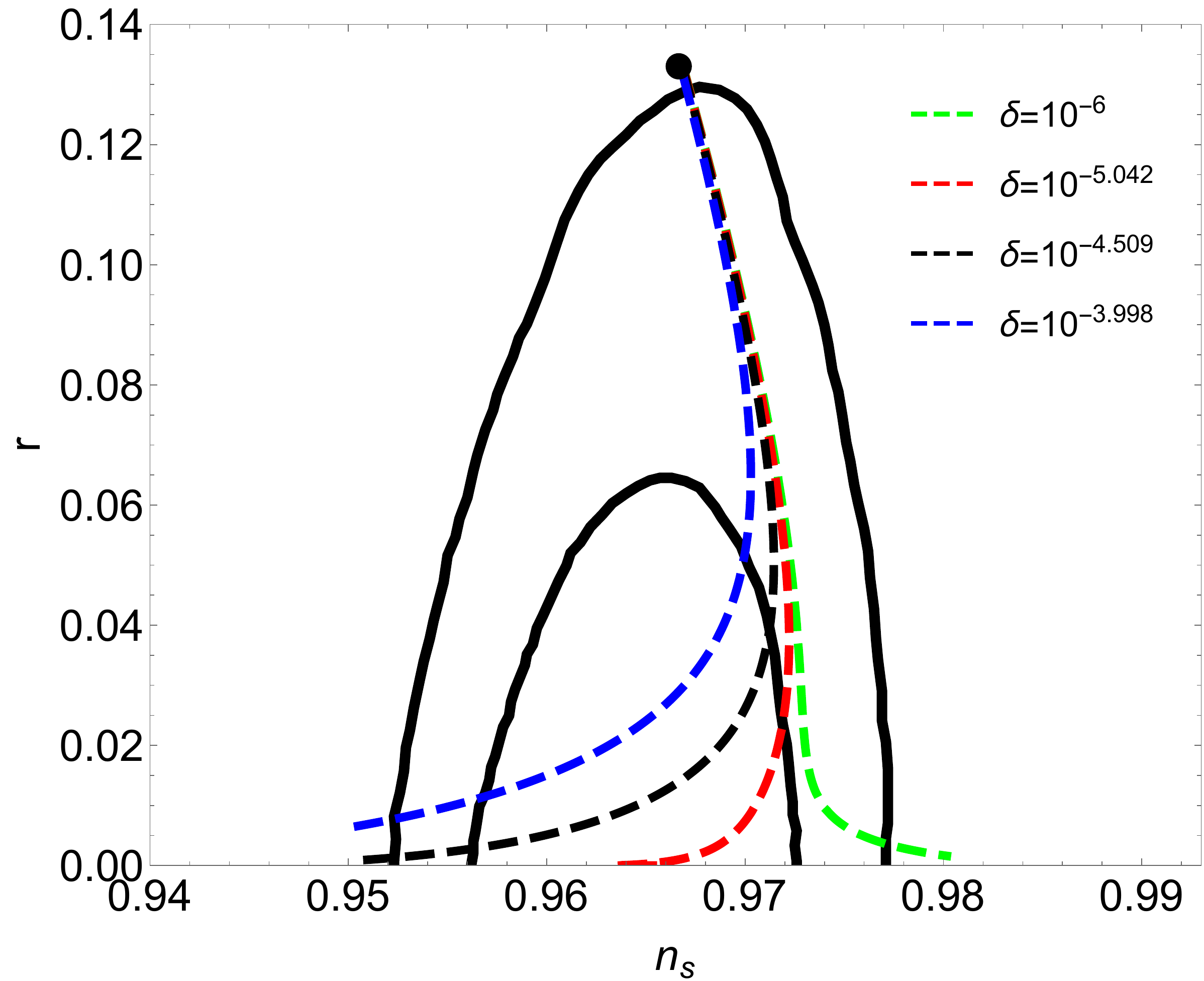}
 		\caption{\small{$n_{s}$ vs r for fixed values of $\delta$}}
 		\label{ns_r_fixed_delta}
 	\end{subfigure}	
 	\begin{subfigure}{.5\textwidth}
 		\centering
 		\includegraphics[width=.9\linewidth]{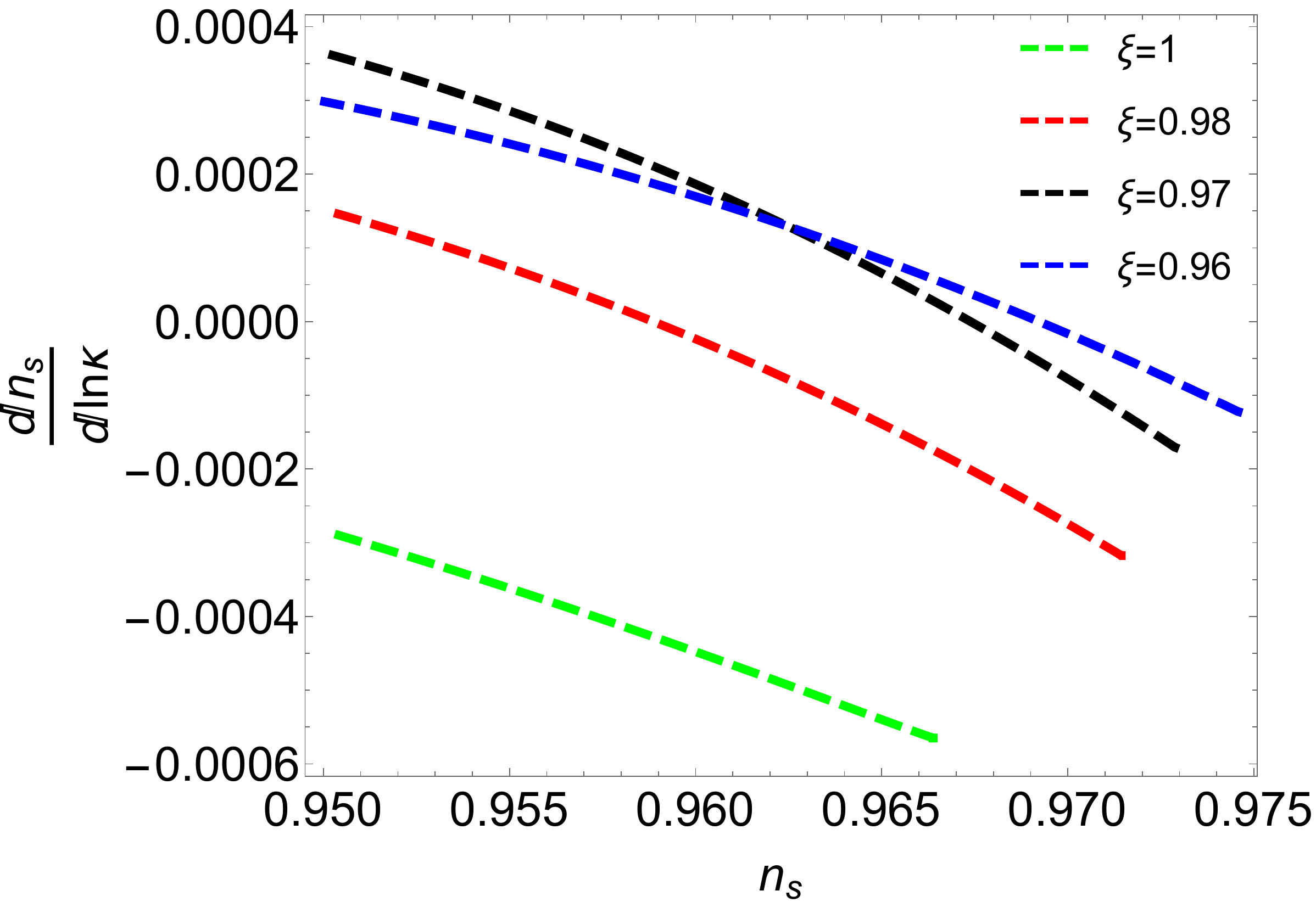}
 		\caption{\small{$\frac{dn_{S}}{d\ln{k}}$ vs $n_{s}$ for fixed values of $\xi$}}
 		\label{ns_dns_fixed_xi}
 	\end{subfigure}%
 	\begin{subfigure}{.5\textwidth}
 		\centering
 		\includegraphics[width=.9\linewidth]{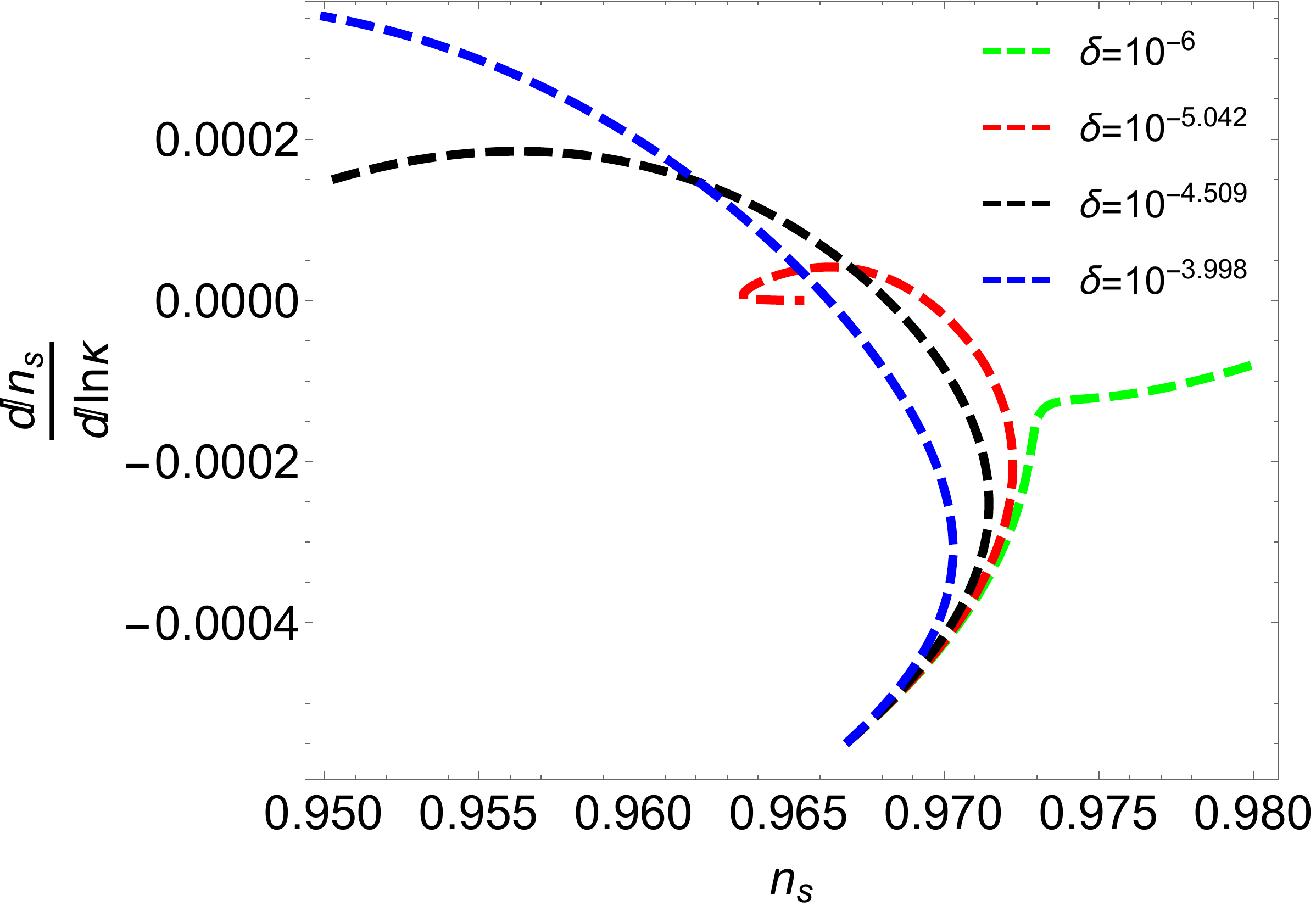}
 		\caption{\small{$\frac{dn_{S}}{d\ln{k}}$ vs $n_{s}$ for fixed values of $\delta$}}
 		\label{ns_dns_fixed_delta}
 	\end{subfigure}%
 	\caption{\small{The inflationary predictions ($r$-$n_{s}$) and ($\frac{dn_{s}}{d\ln{k}}-n_{s}$) of the model by varying the various parameters involved in to the analysis. In all cases we took the  number of e-folds, $N=60$. In plots (a) and (b) black solid contours represents the Planck constraints  (\emph{Planck} TT,TE,EE+lowP)  at $68\%$ (inner) and $95\%$ (outer) confidence level \cite{Ade:2015lrj}. In plots (a) and (c) we keep $\xi$ constant for each curve and vary $\tilde{\lambda}$ and $\delta$. While in plots (b) and (d) for each curve we fixed $\delta$ and vary $\tilde{\lambda}$ and $\xi$. The black dot solution corresponds to $\xi=1$. }}
 	\label{ns_vs_r_plots}
 \end{figure}

   Focusing now on the  numerical analysis, we see that we have to deal with three parameters: $\xi, \delta$ and $\tilde{\lambda}$. We took the number of e-folds ($N$) to be 60, and  in Figure \ref{ns_vs_r_plots} we present two different cases in the $n_{s}-r$ plane, along with the Planck measurements (\emph{Planck} TT,TE,EE+lowP)  \cite{Ade:2015lrj}. Specifically, in Figure $1(a)$, we fixed  $\xi$ and vary $\tilde{\lambda}$ and $\delta$. The various colored (dashed) lines corresponds to different fixed $\xi$-values. The green line corresponds to the limiting case with $\xi=1$ and as we observe the results are more consistent with the Plank bounds (black solid contours) as the value of $\xi$ decreases. Similar, in Figure $1(b)$ we treat $\delta$ as a fixed parameter while we vary $\xi$ and $\tilde{\lambda}$. Also, in this case,  we observe that for a significant region of the parameter space the solutions are in good agreement with the observed cosmological bounds. The green curve here corresponds to $\delta=10^{-6}$. The special case with $\delta=10^{-6}\sim 0$ and $\xi=1$ is represented by the black dot and as we discussed earlier is ruled out from the recent cosmological bounds. We observe from the plot that, as $\xi$ approaches to unity the splitting between the curves due to different values of $\delta$ is small and the solution converges to $\delta\sim{0}$ case. However, as we decrease the values of $\xi$  we have splitting of the curves and better agreement with the cosmological bounds.  Finally in plots 1(c) and 1(d) we present values of the running of the spectral index with respect to $n_{s}$. We observe that the running of the spectral index, approximately receives values in the range  $-5\times{10^{-4}}<\frac{dn_{S}}{d\ln{k}} <5\times{10^{-4}}$.

 Next we present additional plots to better clarify the r\^ole of the various parameters involved in the analysis.
 
\begin{figure}[t!]
	\begin{subfigure}{.5\textwidth}
		\centering
	\includegraphics[width=.9\linewidth]{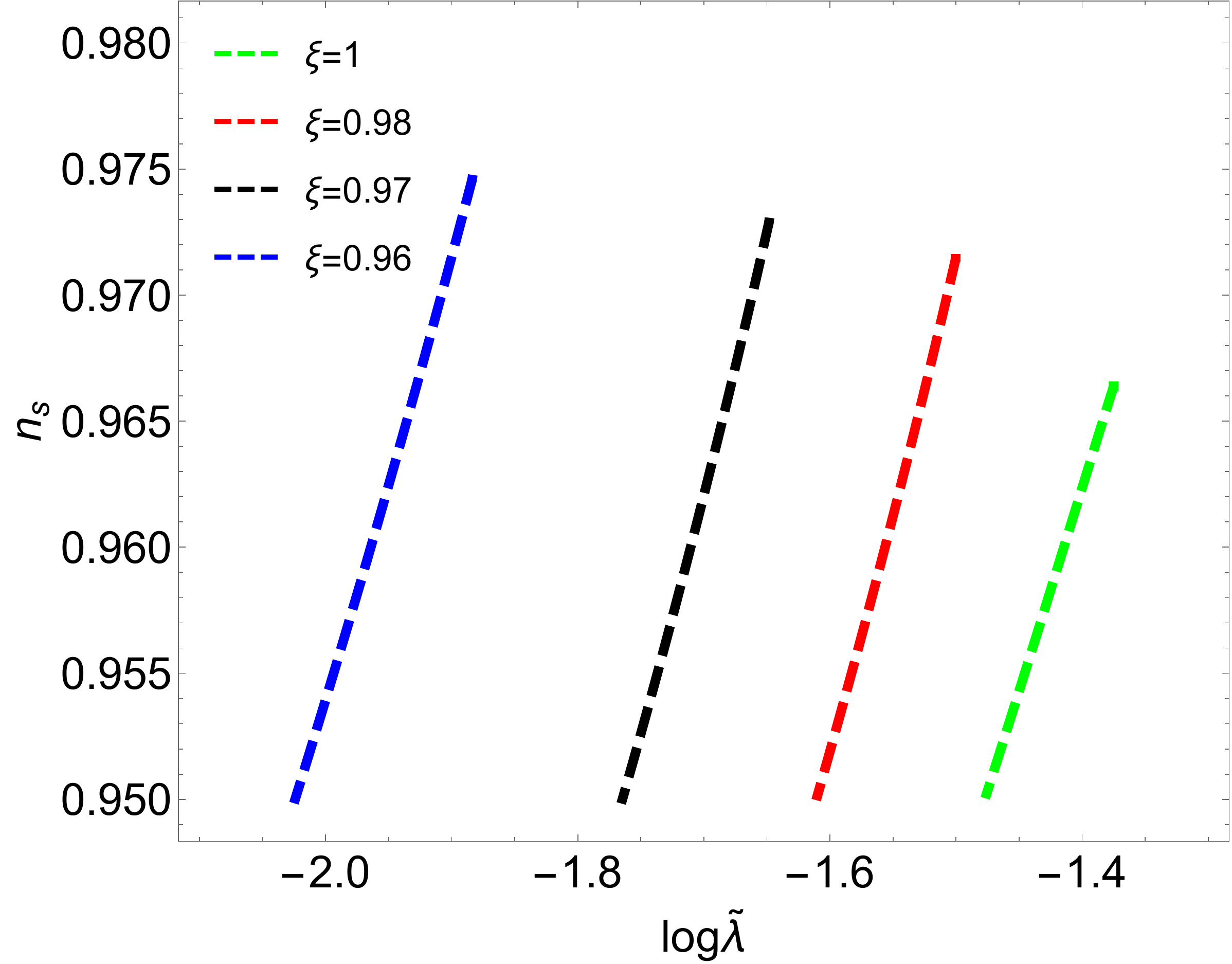}
		\caption{\small{$n_{S}$ vs $\log{\tilde{\lambda}}$}}
		\label{ns_lam_fixed_xi}
	\end{subfigure}%
	\begin{subfigure}{.5\textwidth}
		\centering
	\includegraphics[width=.9\linewidth]{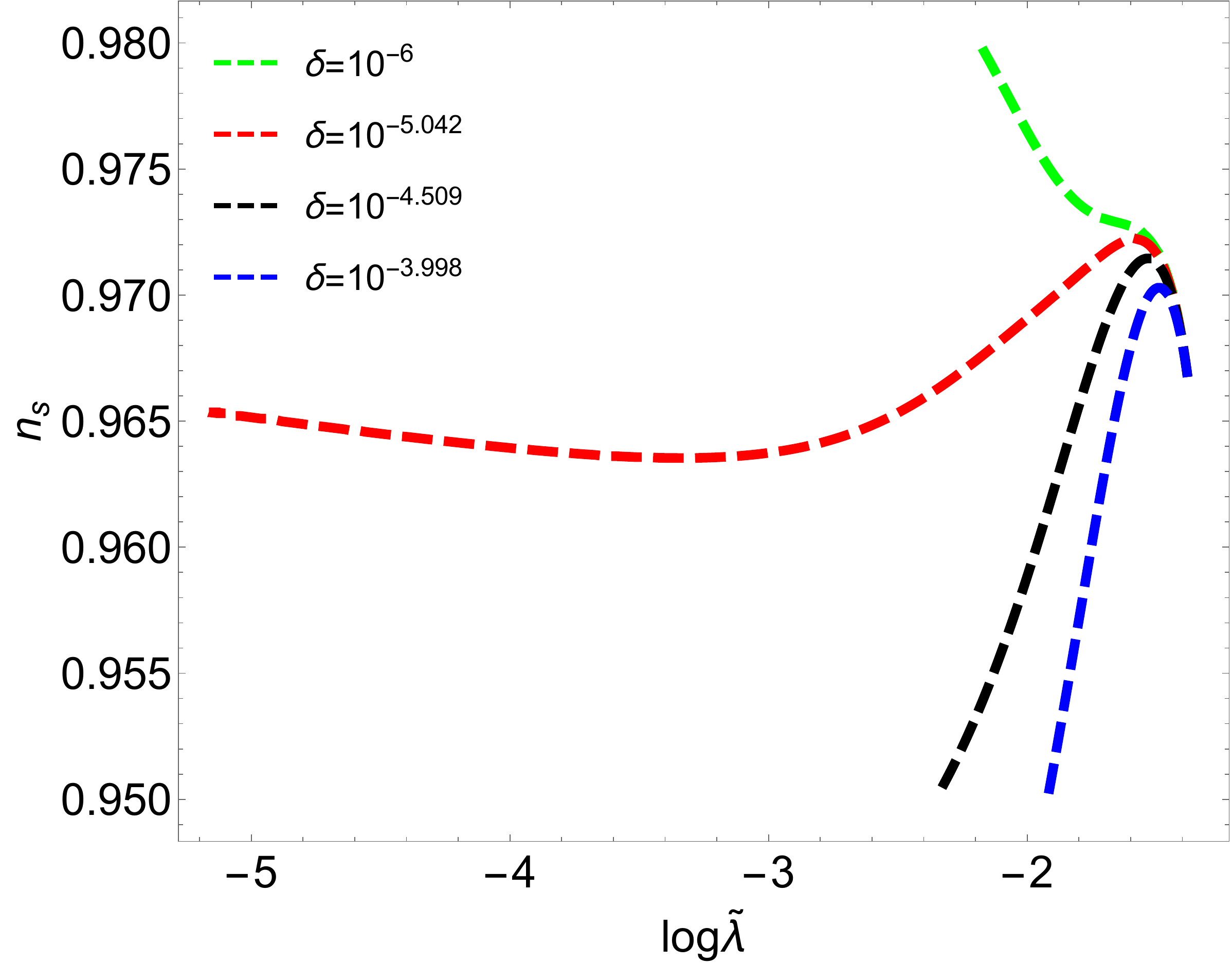}
		\caption{\small{$n_{S}$ vs $\log{\tilde{\lambda}}$}}
		\label{ns_lam_fixed_delta}
	\end{subfigure}
		\medspace\\
	\begin{subfigure}{.5\textwidth}
		\centering
	\includegraphics[width=.9\linewidth]{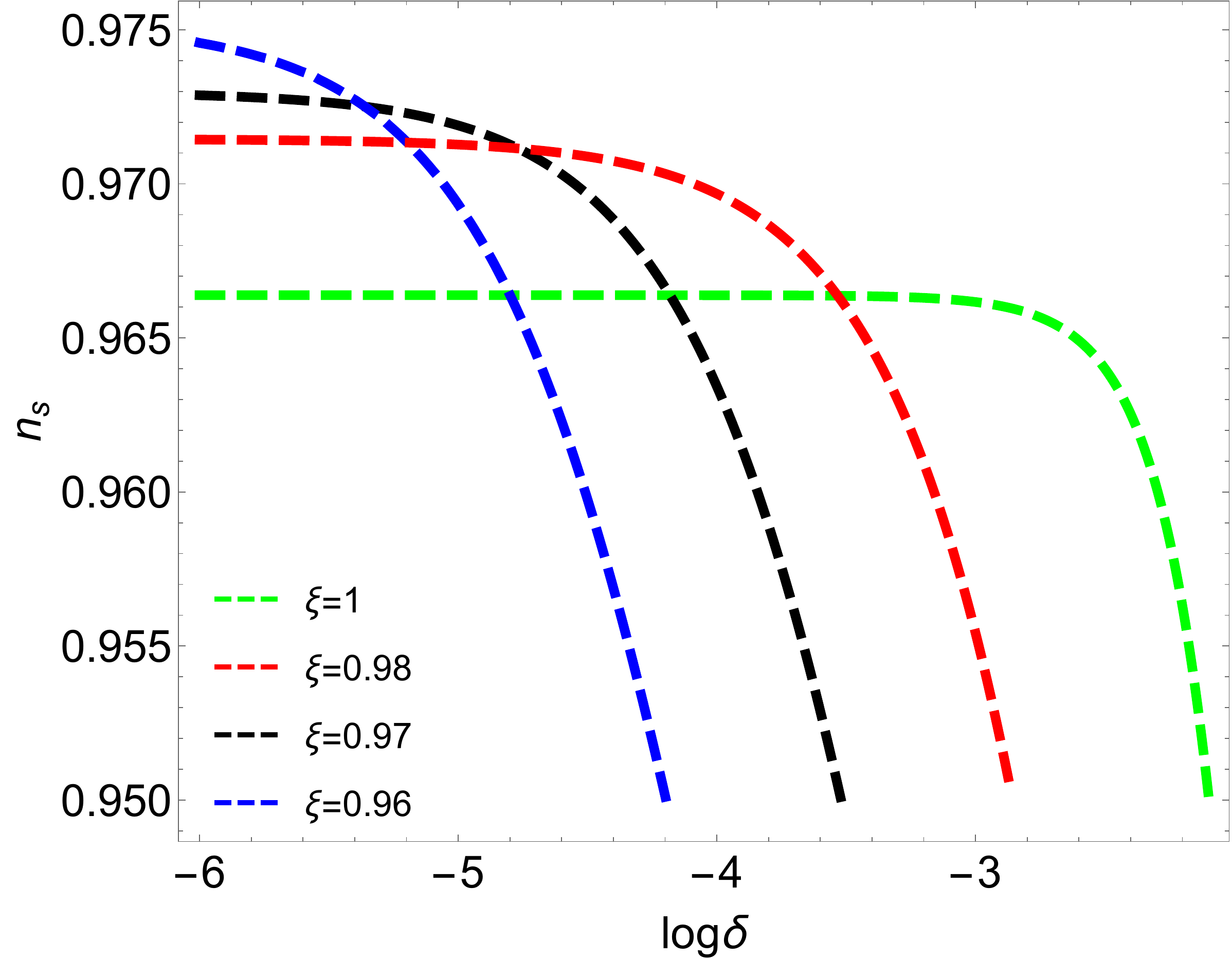}
		\caption{\small{$n_{S}$ vs $\log\delta$}}
		\label{ns_delta}
	\end{subfigure}%
	\begin{subfigure}{.5\textwidth}
		\centering
		\includegraphics[width=.9\linewidth]{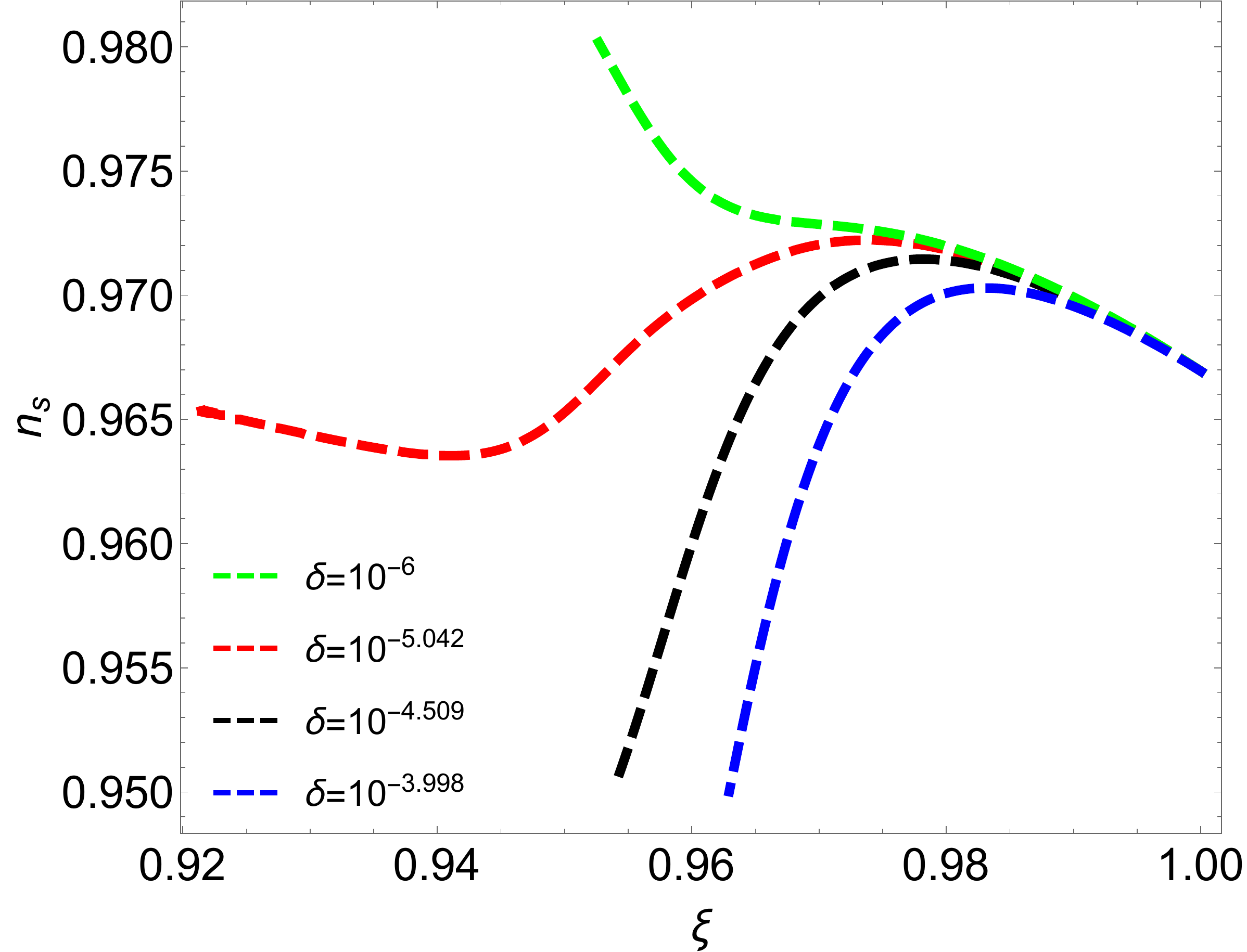}
		\caption{\small{$n_{S}$ vs $\xi$}}
		\label{ns_xi}
	\end{subfigure}%
	\caption{\small{Plots (a) and (c) shows how $n_{S} $ depends on $\log{\tilde{\lambda}}$ and  $\log{\delta}$ respectively. For each curve in Plots (a),(c) we fixed the value of $\xi$ and vary $\tilde{\lambda}$ and $\delta$. Similarly, Plots (b) and (d), shows $n_{s} $ vs $\log{(\tilde{\lambda})}$ and $n_{S} $ vs $\xi$ respectively. In Plots (b) and (d) the value of $\delta$ is fixed while we vary the other parameters.}}
	\label{ns_plots}
\end{figure}

 Firstly, we study the spectral index $n_{s}$ as a function of the various parameters. The results are presented in Figure \ref{ns_plots}. In plots (a) and (b) we consider the cases with fixed values for $\xi$ and $\delta$ respectively, and we take variations for  $\tilde{\lambda}$.  We vary the parameter $\xi$ 
  in the range $\xi\sim{[0.92,1]}$ with the most preferable solutions for $\xi\simeq[0.96, 1]$. In addition the two plots suggest that acceptable solutions 
  are found in the range $\tilde{\lambda}\sim[10^{-2},10^{-1}]$.  In plots (c) and (d)  $n_s$ is depicted  in terms of $\delta$ and $\xi$ respectively. As we expected the dependence on $\delta$ is negligible when it receives very small values, since we observe from plot 3(c) that the various curves are almost constant for very small $\delta$ values. The results are become more sensitive on $\delta$ as we decrease the value of $\xi$. This behaviour can also be confirmed  from the potential \eqref{potentiladelta}. As we can see for $\xi\sim{1}$ the second term is simplified and the potential receives a chaotic like form. In this case the effects of small $\delta$ in the observables are almost negligible (green line). However as we decrease the value of $\xi$ and we increase the values of $\delta$ the second term becomes important and contributes to the results.

\noindent

\begin{figure}[t!]
 	\begin{subfigure}{.5\textwidth}
 		\centering
 	\includegraphics[width=.95\linewidth]{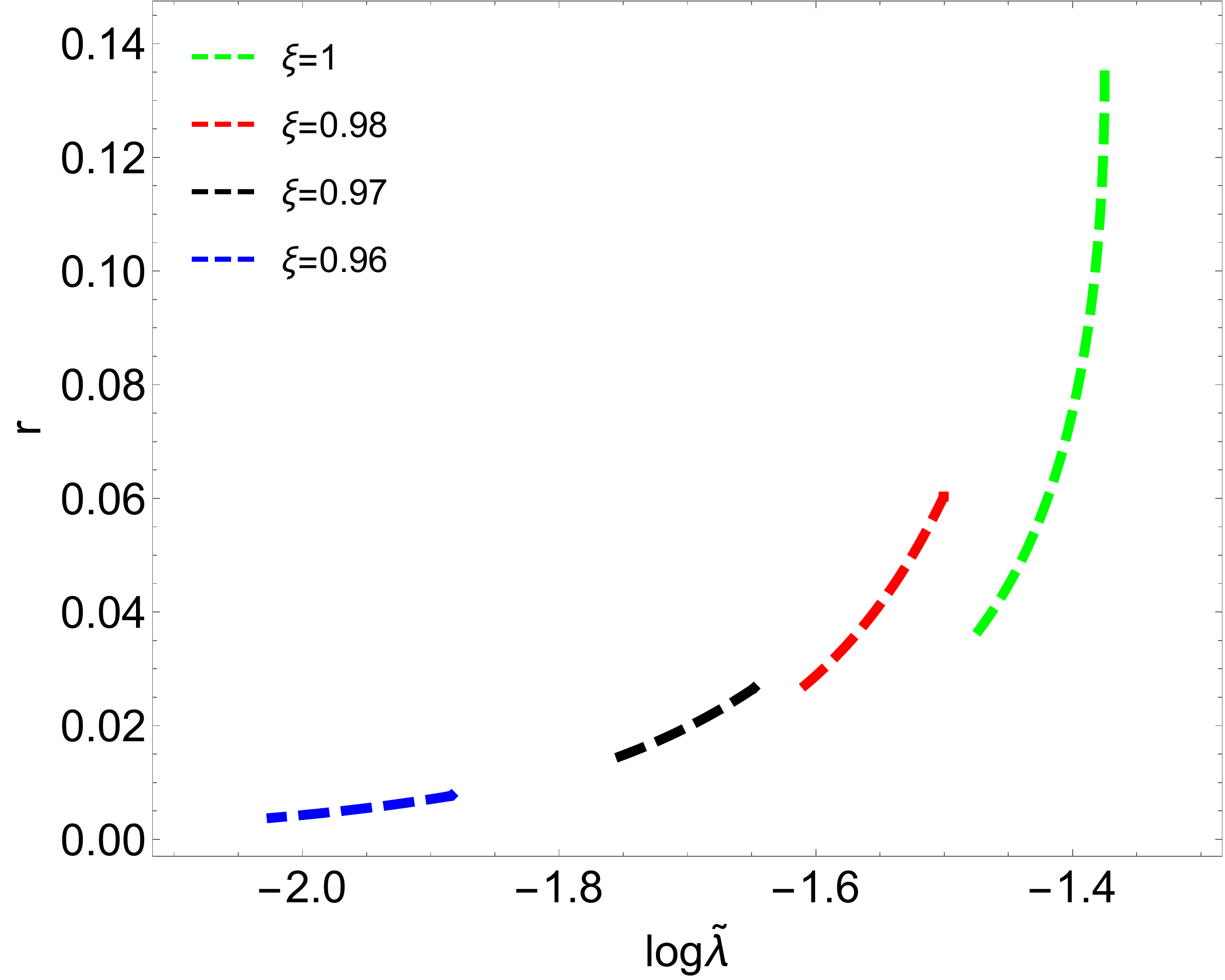}
 	\caption{ $r$ vs $\log{\tilde{\lambda}}$}
 	\label{r_lam_fixed_xi}
 	\end{subfigure}%
 	\begin{subfigure}{.5\textwidth}
 		\centering
 	\includegraphics[width=.95\linewidth]{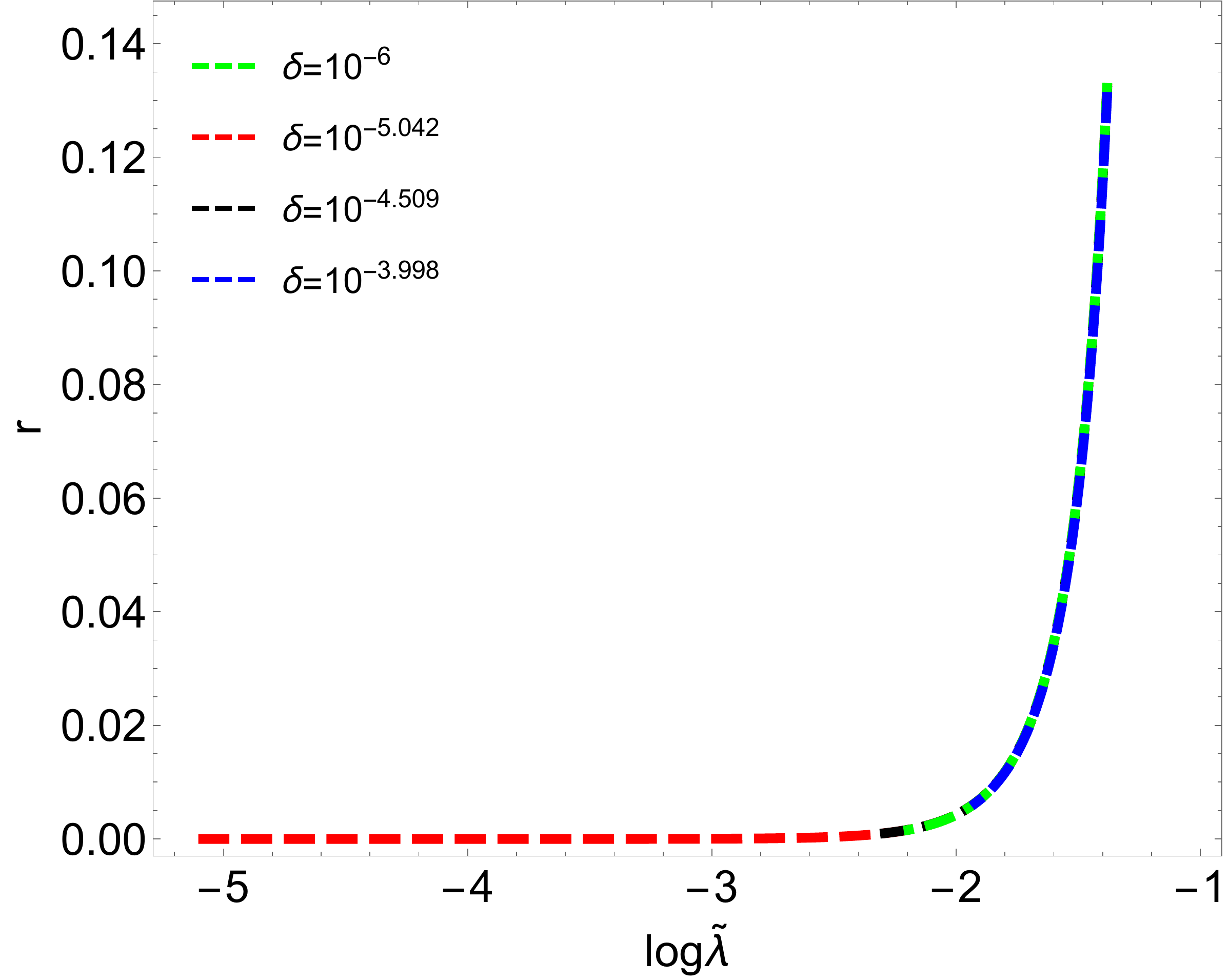}
 	\caption{ $r$ vs $\log{\tilde{\lambda}}$}
 	\label{r_lam_fixed_delta}
 	\end{subfigure}%
 	\medspace\\	
 	 	\begin{subfigure}{.5\textwidth}
 	 		\centering
 	 		\includegraphics[width=.95\linewidth]{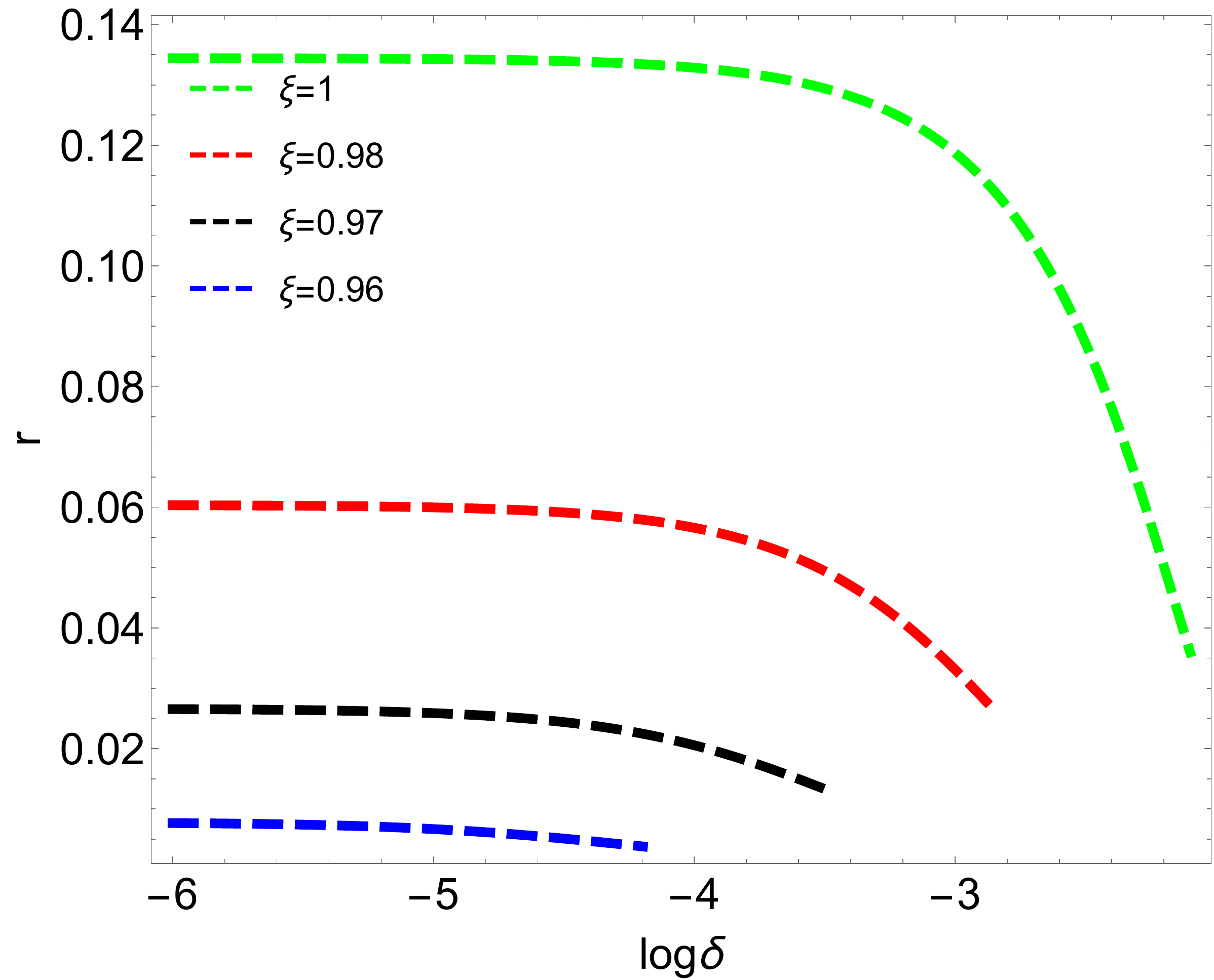}
 	 		\caption{ $r$ vs $\log\delta$}
 	 		\label{r_delta}
 	 	\end{subfigure}%
 	 	\begin{subfigure}{.5\textwidth}
 	 		\centering
 	 		\includegraphics[width=.95\linewidth]{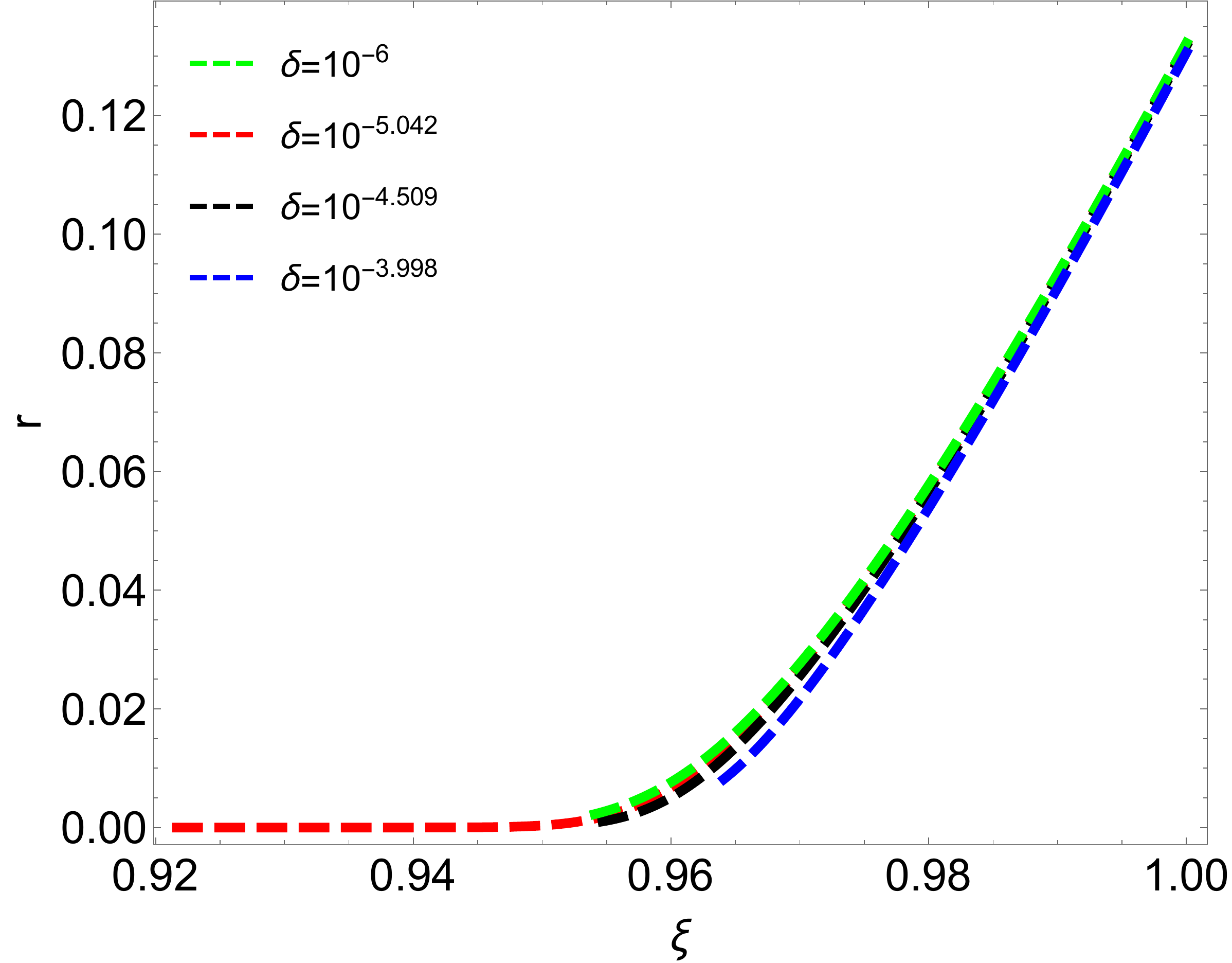}
 	 		\caption{ $r$ vs $\xi$}
 	 		\label{r_xi}
 	 	\end{subfigure}
 	\caption{\small{Plots (a) and (c) shows $r$ vs $\log{\tilde{\lambda}}$ and $r$ vs $\log{\delta}$ respectively. For each curve in Plots (a) and (c) we fixed $\xi$  and vary $\tilde{\lambda}$ and $\delta$. Similar, in Plots (b) and (d) we present  $r$ vs $\log{\tilde{\lambda}}$ and $r$ vs $\xi$. For each curve in these plots we fixed the value of $\delta$ and vary $\tilde{\lambda}$ and $\xi$.}}
 	\label{r_plots}
 \end{figure}

Next, in Figure \ref{r_plots} we consider various cases for the tensor to scalar ratio, r. The description of the plots follows the spirit of those presented in Figure \ref{ns_plots} for  the spectral index $n_{S}$. In particular, by comparing the plots 4(c) and 3(c) we notice that the dependence of $r$ on $\delta$ is weaker in comparison with $n_{S}$. Thus the relaxation parameter $\delta$ strongly affects the spectral index $n_{S}$ while for $\delta<10^{-4}$ and fixed $\xi$ the tensor-scalar ratio $r$ remains almost constant. In summary from the various figures presented so far we observe that  consistent solutions can be found in a wide range of the parameter space. We also note that the model predicts solutions with $r\leq{0.02}$, which is a prediction that can be tested with the discovery of primordial gravity waves and with bounds of future experiments. 

Regarding the superpotential parameter $\tilde{\lambda}$, we can see from the various plots that its value must be  within the range $\tilde{\lambda}\sim{[10^{-2}, 10^{-1}}]$. Using this range of values for $\tilde{\lambda}$ and the fact that, $M_{Q_{H}}\approx{\frac{8\tilde{\lambda}}{9}\upsilon^{2}}$, with $\upsilon\simeq{10^{-2}}$ in $M_{Pl}=1$ units we conclude that : $M_{Q_{H}}\sim{[0.217, 2.17]\times{10^{13}}}$ GeV. The fact that the mass value is small compare to the $\mathcal{O}(M_{GUT})$ scale, can create tension with other phenomenological predictions of the model, like unification of gauge couplings. On the other hand, as already mentioned  , $Q_{H},\ov{Q}_{H}$ triplet fields can be mixed with the triplets   $D_{3},\ov{D}_{3}$ contained in the sextet $D_{6}$, something that is possible to lead in a significant lift to the mass value of the extra triplet fields.

\begin{figure}[t!]
 	\begin{subfigure}{.5\textwidth}
 		\centering
 		\includegraphics[width=.9\linewidth]{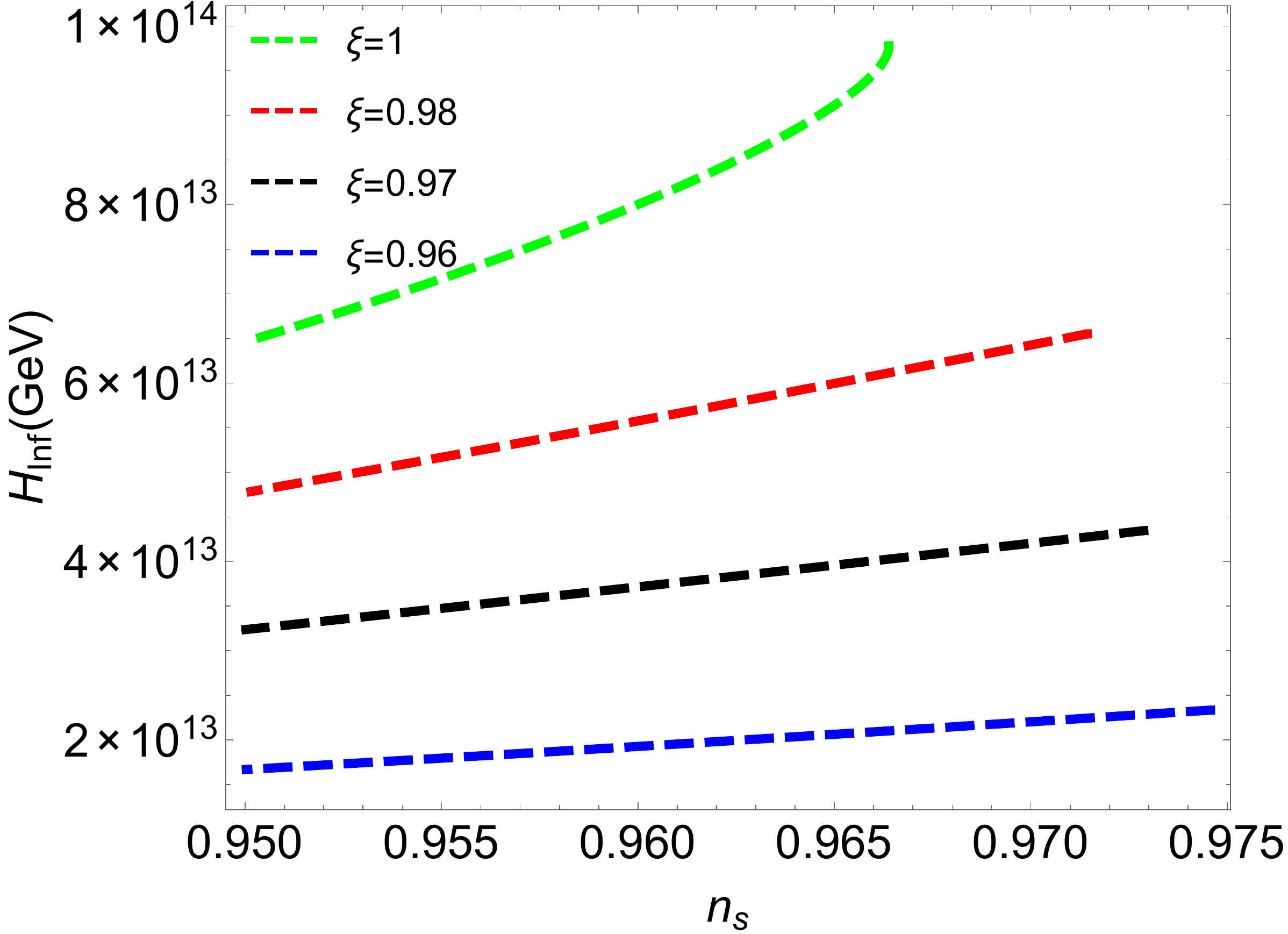}
 		\caption{\small{$H_{inf}$ vs $n_{s}$ for fixed values of $\xi$}}
 		\label{Hinf_ns_fixed_xi}
 	\end{subfigure}%
 	\begin{subfigure}{.5\textwidth}
 		\centering
 		\includegraphics[width=.9\linewidth]{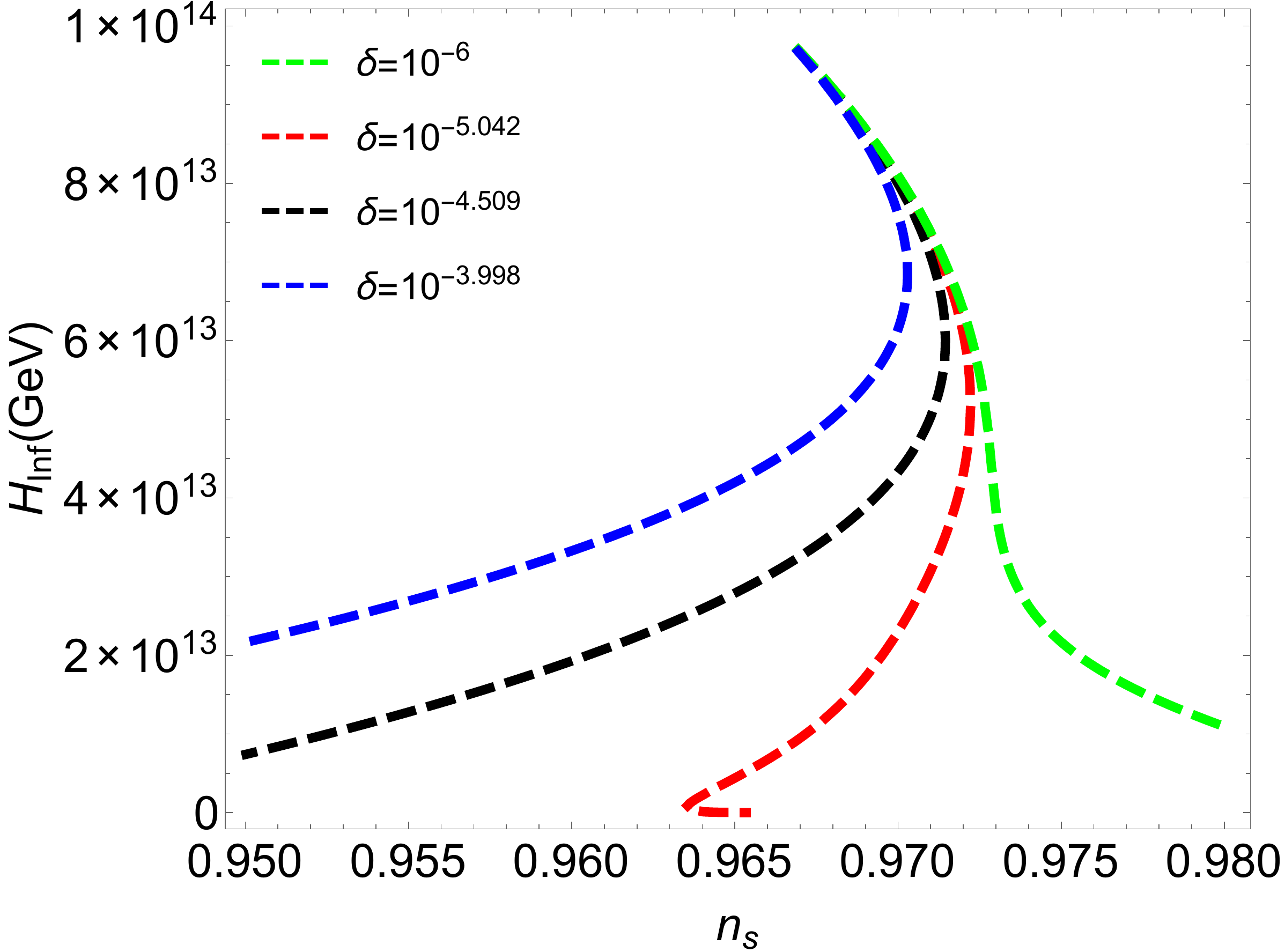}
 		\caption{\small{$H_{inf}$ vs $n_{s}$ for fixed values of $\delta$}}
 		\label{ns_r_fixed_delta}
 	\end{subfigure}	
 	\caption{\small{Plots showing the values (in GeV) of the Hubble parameter with respect to the scalar spectral index $n_{s}$. For acceptable $n_{s}$ values we see that the Hubble parameter receives values of order~$10^{13}-10^{14}$ GeV.}}
 	\label{Hinf_ns_plots}
 \end{figure}

It is also interesting to investigate the values of the Hubble parameter during inflation $H_{inf}$ in the model. In the slow-roll limit the Hubble parameter it depends on the value of $X$:

\begin{equation}
H_{inf}^{2}=\frac{V(X)}{3M_{Pl}^{2}}
\end{equation}

\noindent and we evaluate it at the pivot scale. In Figure \ref{Hinf_ns_plots} we show the values of the Hubble parameter in the ($H_{inf}-n_{s}$) plane. We observe that the values of the Hubble parameter with respect to $n_{s}$ bounds are of order $10^{13}$ GeV.

\subsection{REHEATING}

As already have been discussed in Section 2, the quarks and leptons in the 4-2-2 model are unified under the representations $F_{i}=(4,2,1)$ and $\bar{F}_{i}=(\bar{4},1,2)$, where $i=1,2,3$ denote the families and the RH-neutrinos are contained in the $\bar{F}$ representation. A heavy Majorana mass for the RH-neutrinos can be realized from the following non-renormalisable term 

\be \label{majorana}
M_{\nu^c} \nu^c\nu^c\approx 
  \gamma\frac{\bar{F}\bar{F}\bar{H}\bar{H}}{M_{*}}
  \ee

\noindent where we have suppressed generation indices for simplicity, $\gamma$ is a coupling constant and $M_{*}$ represents a high cut-off scale (for example the compactification scale in a string model or the Planck scale $M_{Pl}$). In terms of $SO(10)$  GUTs this operator descent from the following invariant operator

\[ 16_{F}16_{F}\bar{16}_{H}\bar{16}_{H}\] 

  \noindent and as described in \cite{Leontaris:2016jty} can be used to explain the reheating process of the universe after the end of inflation. In our case the 4-2-2 symmetry breaking occur in two steps: first $G_{PS}\xrightarrow{\langle{S}\rangle}G_{L-R}$ and then  $G_{L-R}\xrightarrow{\langle{\nu_{H}}\rangle,\langle{\bar{\nu}_{H}}\rangle}G_{SM}$. The first breaking is achieved via the adjoint of the PS group at the GUT scale while the second breaking occurs in an intermediate scale $M_{R}$. After the breaking of the L-R symmetry, the high order term in (\ref{majorana}) gives the following Majorana mass term for the RH neutrinos
  
\be  
  \gamma\frac{\langle{\nu_{H}}\rangle^{2}}{M_Pl}\nu^{c}\nu^{c}.
  \ee

\begin{figure}[t!]
	\begin{subfigure}{.5\textwidth}
		\centering
		\includegraphics[width=.9\linewidth]{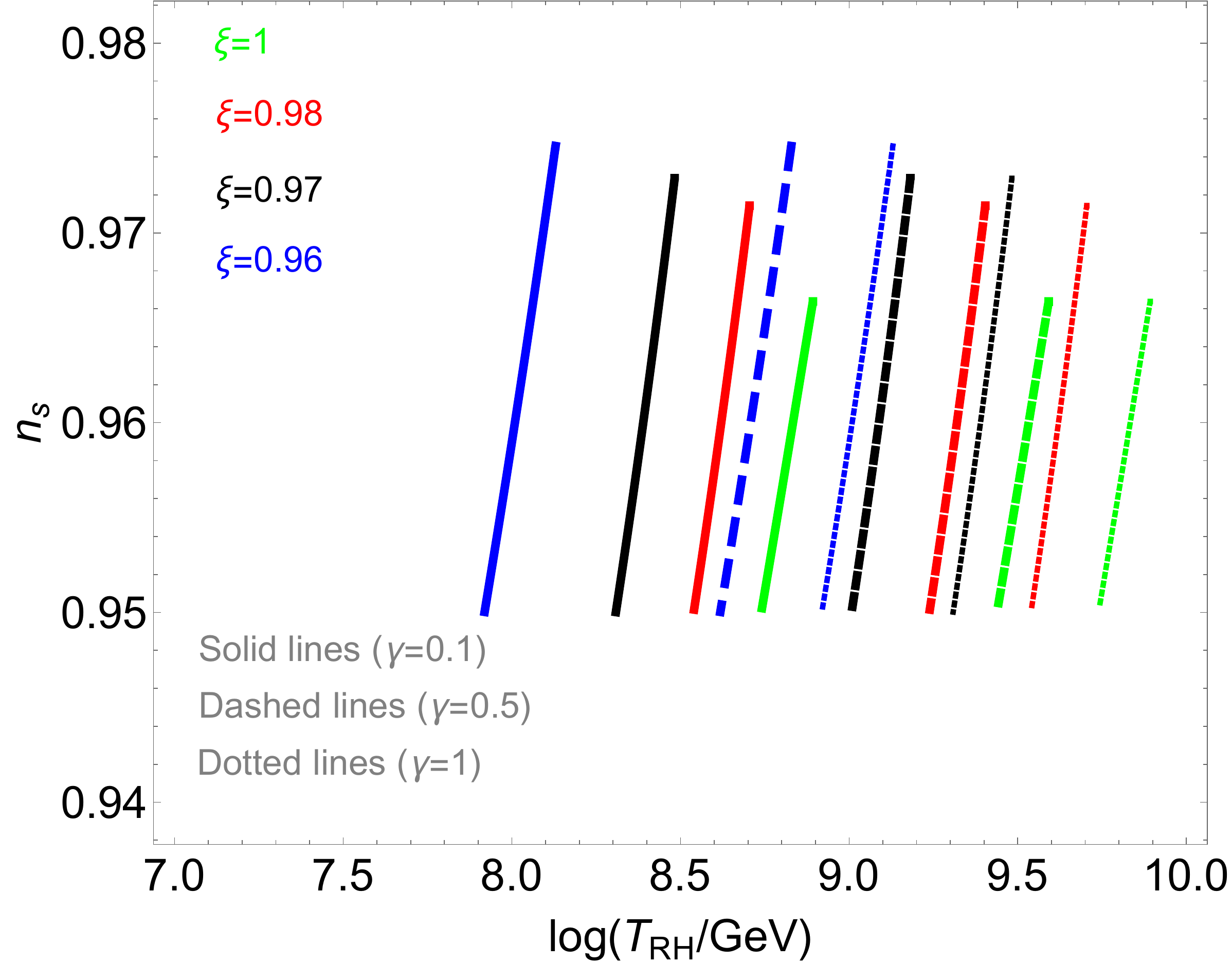}
		\caption{}
		\label{ns_Trh_fixed_xi01}
	\end{subfigure}%
	\begin{subfigure}{.5\textwidth}
		\centering
		\includegraphics[width=.9\linewidth]{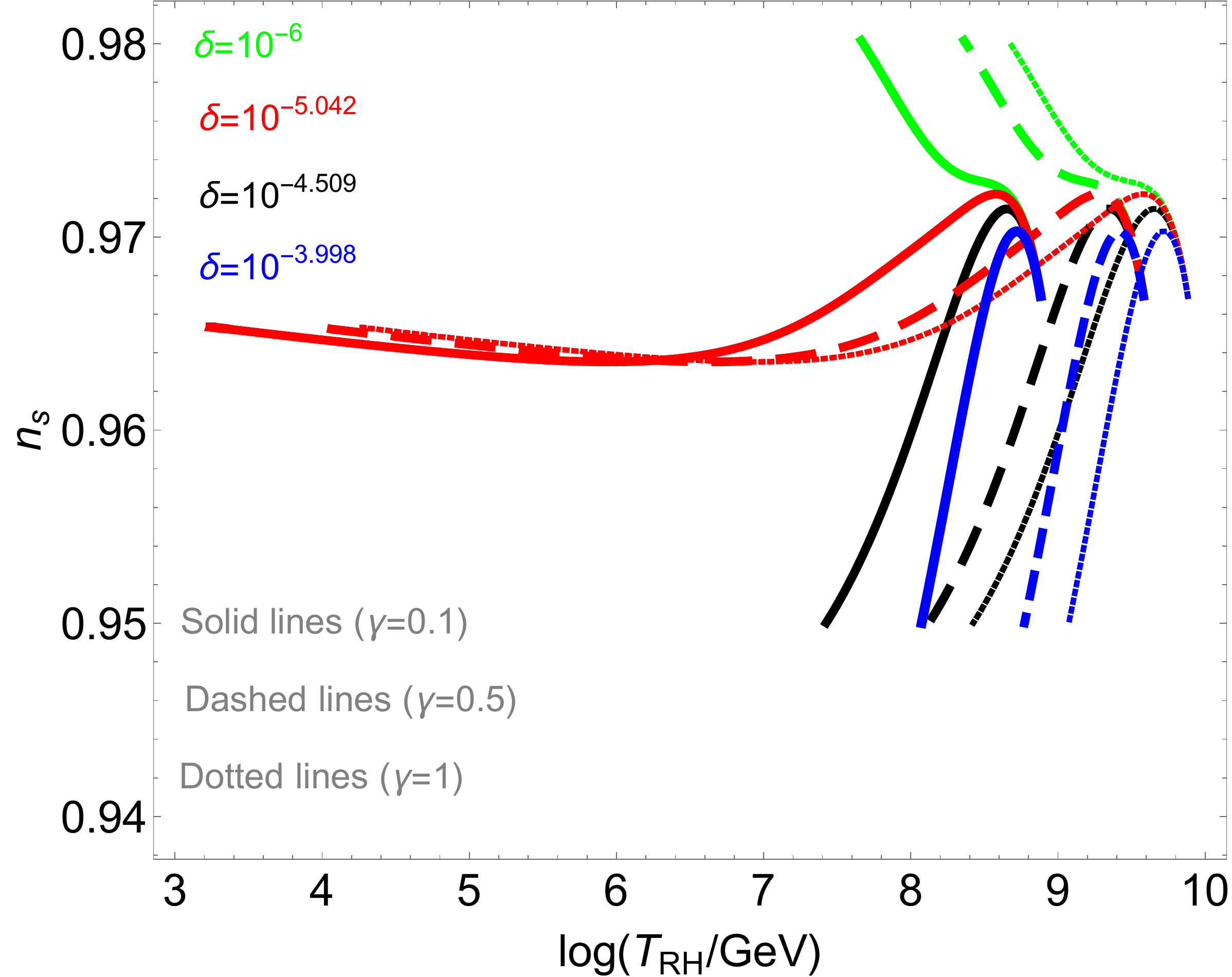}
		\caption{ }
		\label{ns_Trh_fixed_xi05}
	\end{subfigure}%
	\medspace\\	
	\begin{subfigure}{.5\textwidth}
		\centering
		\includegraphics[width=.9\linewidth]{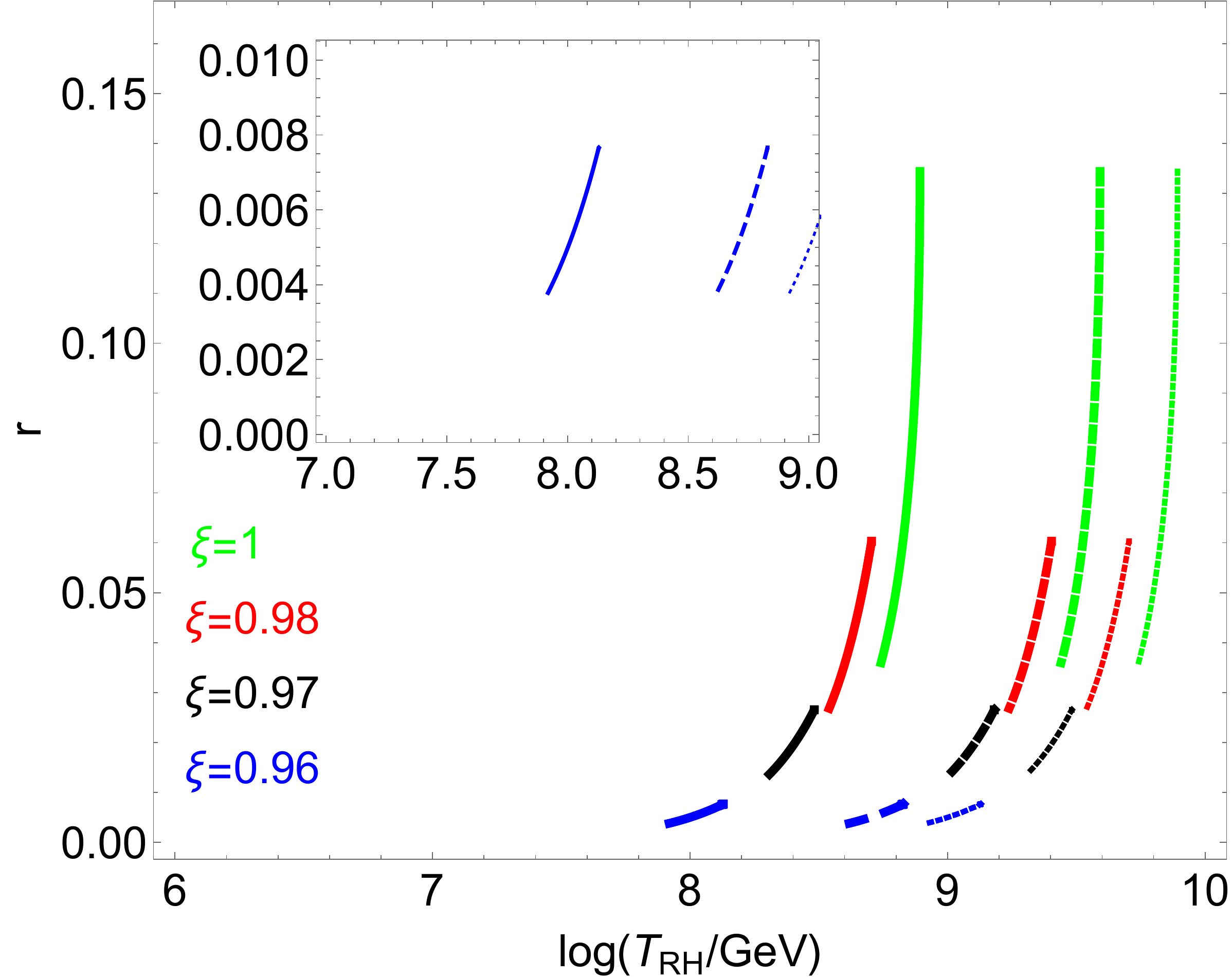}
		\caption{}
		\label{r_Trh_fixed_xi}
	\end{subfigure}%
	\begin{subfigure}{.5\textwidth}
		\centering
		\includegraphics[width=.9\linewidth]{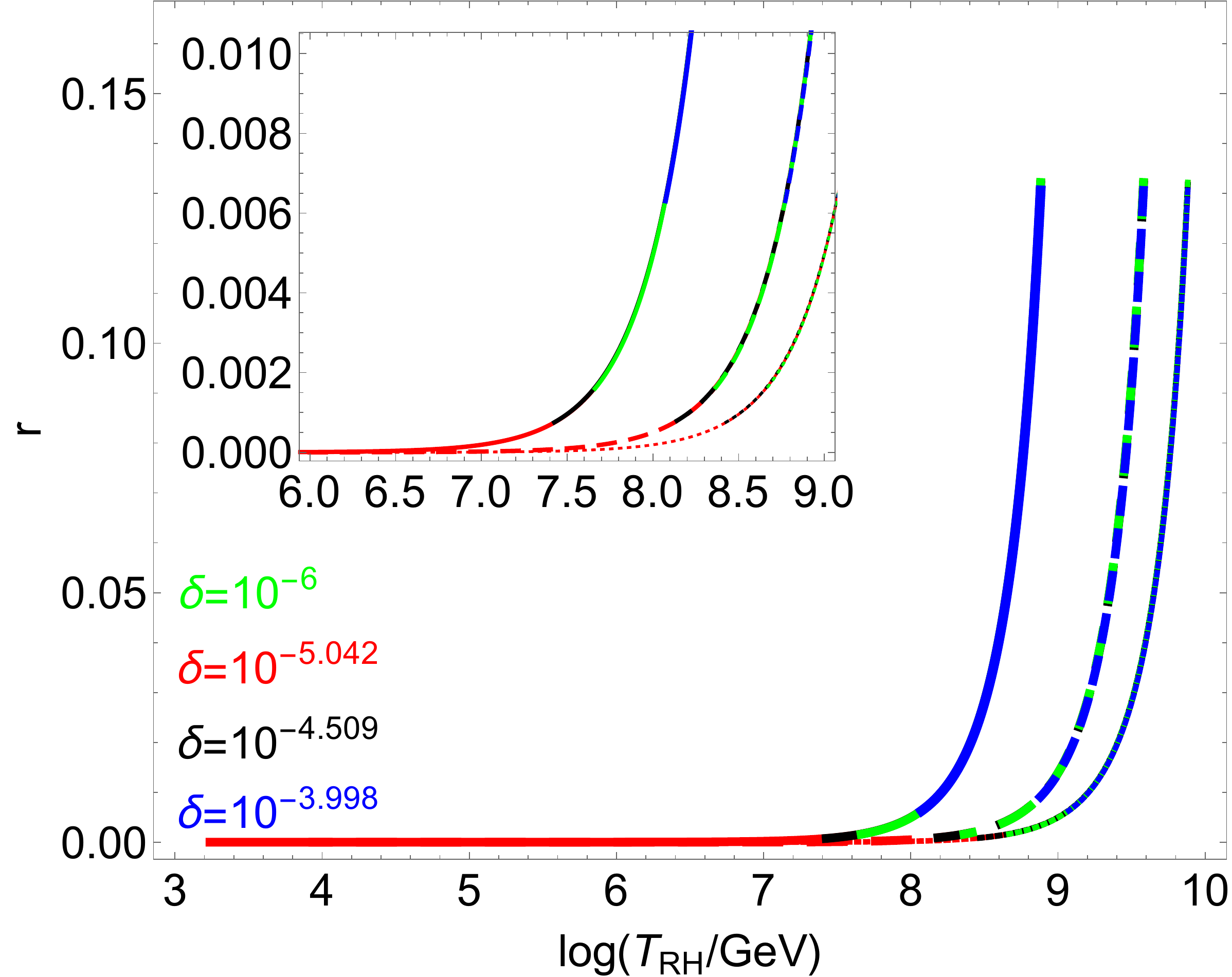}
		\caption{ }
		\label{r_Trh_fixed_delta}
	\end{subfigure}
	\caption{\small{Plots (a) and (b) shows solutions in the $n_{s}-T_{RH}$ plane by varying the various parameters of the model, while plots (c) and (d) present solutions in the $r-T_{RH}$ plane. In all the cases for the coupling constant $\gamma$ we choose the values $\gamma=0.1$ (solid), $\gamma=0.5$ (dashed) and $\gamma=1$ (dotted).}}
	\label{ns_Trh_plots}
\end{figure}

\noindent We can see that a heavy Majorana scale scenario implies that the $SU(2)_{R}$ breaking scale should not be much lower than the $SU(4)$ scale and also $\gamma$ should not be too small. Another important role of the higher dimensional operators is that after inflation the
inflaton $X$ decays into RH neutrinos through them to reheat the Universe. In addition the subsequent decay of these neutrinos can explain the baryon asymmetry via leptogenesis \cite{Fukugita:1986hr, Lazarides:1991wu}
. For the reheating temperature, we  estimate \cite{Leontaris:2016jty} (see also \cite{Lazarides:2001zd}) :

\begin{equation}
T_{RH}\sim \sqrt{\Gamma_{X} M_{Pl}}
\end{equation}

 \noindent where the total decay width of the inflaton is given by

 \be
\Gamma_{X}\simeq{\frac{1}{16\pi}\left(\frac{M_{\nu^{c}}}{M}\right)^{2}M_{X}}  
\ee

\noindent with $M_{\nu^{c}}= \gamma\frac{\langle{\nu_{H}}\rangle^{2}}{M_{Pl}}$ the mass of the RH neutrinos and $M_{X}$ the mass of the inflaton. The later is calculated from the effective mass matrix at the local minimum and approximately is   $M_{X}=2M\simeq{2\tilde{\lambda }\upsilon^{2}}$. Since $M\simeq{10^{13}}$GeV, the decay condition $M_{X}>M_{\nu^{c}}$ it is always satisfied for appropriate choices of the parameters $\langle\nu_{H}\rangle$ and $\gamma$. In Figures \ref{ns_Trh_plots} we present solutions in $n_{s}-T_{RH}$ and $r-T_{RH}$ plane with respect to the various parameters of the model. For the computation of $T_{RH}$ we assume that $\langle\nu_{H}\rangle =M\simeq{\tilde{\lambda}v^{2}}$ and we present the results for  $\gamma=0.1$ (solid), $\gamma=0.5$ (dashed) and $\gamma=1$ (dotted). In this range of $\gamma$ values we have a Majorana mass, $M_{\nu^{c}}\sim{10^{6}-10^{7}}$ GeV, which decreases as we decrease the value of $\gamma$. In addition, gravitino constraints implies a bound for the reheating temperature with $T_{RH}<10^{6}-10^{9}$ GeV and as we observe from the plots there are acceptable solutions in this range of values. More precisely, from plots (a) and (c) we see that for $\xi>0.97$  and $\gamma>0.5$ most of the results predict $T_{RH}>10^{9}$ GeV. However, it is clear that the consistency with the gravitino constraints strongly improves as we decrease $\gamma$, since all the curves with $\gamma=0.1$ (solid lines) predicts $T_{RH}\lesssim{10^{9}}$ GeV. Similar conclusions can be derived from plots (b) and (d). In addition, from the $r-T_{RH}$ plots (c) and (d) we observe that for  $T_{RH}<10^{6}-10^{9}$ there are regions in the parameter space with $r\sim{10^{-2}-10^{-3}}$. Furthermore, we observe from plot 6(c) that the tensor-scalar ratio and the reheating temperature are decreased as we decrease the value of $\xi$ since the curves are shift to the left and down regions of the plot.

A sample of the results have been discussed so far is presented in Table \ref{mastertable}. The table is organized in horizontal blocks and each block contains three sets of values. For each set in a block we change only the coupling constant $\gamma$ ($\gamma=1,0.5,0.1$) while we keep $\tilde\lambda$, $\xi$ and $\delta$ constant. We observe that as we decrease the values of $\tilde{\lambda}$ and $\xi$ the values of the tensor to scalar ratio ($r$) and the reheating temperature ($T_{RH}$) also decreased.

\begin{table}[t]
	\resizebox{\textwidth}{!}{
	\begin{tabular}{|lc|cccc|cc|ccc|c|}
		\hline

		$\frac{X_{0}}{M_{Pl}}$&$\frac{X_{e}}{M_{Pl}}$&$\gamma$&$\tilde{\lambda}$ & $\xi$ &$ \delta$  & $\frac{M_{Inf}}{M_{Pl}}$ & $\frac{M_{\nu^{c}}}{M_{Pl}}$ & $n_{s}$&$r$&$\frac{dn_{s}}{dln\kappa}$&$\log{(T_{RH}/ GeV)}$\\
		\hline
			15.04 &1.41& 1&0.0384&0.9936& $10^{-6}$& $1.16\times10^{-5}$& $3.4\times10^{-11}$& 0.968& 0.1070& $-4.7\times{10^{-4}}$&9.83\\
			15.04 &1.41& 0.5&0.0384&0.9936& $10^{-6}$& $1.16\times10^{-5}$& $1.7\times10^{-11}$& 0.968& 0.1070& $-4.7\times{10^{-4}}$&9.53\\
			15.04 &1.41& 0.1&0.0384&0.9936&$10^{-6}$& $1.16\times10^{-5}$& $3.4\times10^{-12}$& 0.968& 0.1070& $-4.7\times{10^{-4}}$&8.84\\
		\hline
13.848 &1.41& 1& 0.0304&0.98&  $10^{-4.61}$& $9.25\times10^{-6}$& $2.139\times10^{-11}$& 0.971& 0.057& $-2.87\times{10^{-4}}$&9.683\\
13.848 &1.41& 0.5& 0.0304&0.98&  $10^{-4.61}$& $9.25\times10^{-6}$& $1.07\times10^{-11}$& 0.971& 0.057& $-2.87\times{10^{-4}}$&9.382\\
13.848&1.41& 0.1& 0.0304&0.98&  $10^{-4.61}$& $9.25\times10^{-6}$& $2.139\times10^{-12}$& 0.971& 0.057& $-2.87\times{10^{-4}}$&8.683\\
	\hline

	

		12.83& 1.40&1& 0.02141& 0.97&  $10^{-4.22}$& $6.5\times10^{-6}$&
			$1.05\times10^{-11}$& 0.967& 0.0238&  $1.5\times{10^{-6}}$& 9.45\\
			
			12.83& 1.40&0.5& 0.02141& 0.97&  $10^{-4.22}$& $6.5\times10^{-6}$&
			$5.29\times10^{-12}$& 0.967& 0.0238&  $1.5\times{10^{-6}}$& 9.15\\
			
				12.83& 1.40&0.1& 0.02141& 0.97&  $10^{-4.22}$& $6.5\times10^{-6}$&
				$1.05\times10^{-12}$& 0.967& 0.0238&  $1.5\times{10^{-6}}$& 8.45\\
				\hline
			12.69& 1.40&1& 0.019& 0.97&  $10^{-3.72}$& $5.8\times10^{-6}$&
			$8.4\times10^{-12}$& 0.958& 0.018&  $2.3\times{10^{-4}}$& 9.38\\
			
			12.69& 1.40&0.5& 0.019& 0.97&  $10^{-3.72}$& $5.8\times10^{-6}$&
			$4.2\times10^{-12}$& 0.958& 0.018&  $2.3\times{10^{-4}}$& 9.08\\
			
			12.69& 1.40&0.1& 0.019& 0.97&  $10^{-3.72}$& $5.8\times10^{-6}$&
			$8.4\times10^{-13}$& 0.958& 0.018&  $2.3\times{10^{-4}}$& 8.3\\
			\hline
			11.85& 1.40&1& 0.0118&0.96&  $10^{-4.82}$& $3.57\times10^{-6}$& 
			$3.2\times10^{-12}$& 0.966& 0.0061&  $5.1\times{10^{-5}}$& 9.065\\
			
				11.85& 1.40&0.5& 0.0118&0.96&  $10^{-4.82}$& $3.57\times10^{-6}$& 
				$1.6\times10^{-12}$& 0.966& 0.0061&  $5.1\times{10^{-5}}$& 8.76\\
				
					11.85& 1.40&0.1& 0.0118&0.96&  $10^{-4.82}$& $3.57\times10^{-6}$& 
					$3.2\times10^{-13}$& 0.966& 0.0061&  $5.1\times{10^{-5}}$& 8.065\\
				\hline
			11.79& 1.40& 1&0.010& 0.96 & $10^{-4.397}$& $3.13\times10^{-6}$ &
			$2.5\times10^{-12}$& 0.957& 0.0050&  $2.1\times{10^{-4}}$& 8.98\\
			
				11.79& 1.40&0.5& 0.010& 0.96 & $10^{-4.397}$& $3.13\times10^{-6}$ &
				$1.2\times10^{-12}$& 0.957& 0.0050&  $2.1\times{10^{-4}}$& 8.67\\
			
				11.79& 1.40&0.1& 0.010& 0.96 & $10^{-4.397}$& $3.13\times10^{-6}$ &
				$2.5\times10^{-13}$& 0.957& 0.0050&  $2.1\times{10^{-4}}$& 7.97\\
			
		\hline
			11.64& 1.404&1&0.00891&0.958& ${10^{-4.5}}$& $2.71\times10^{-6}$& 
			$1.85\times10^{-12}$& 0.957& 0.0034&  $1.8\times{10^{-4}}$& 8.89\\
			
			11.64& 1.404&0.5&0.00891&0.958& ${10^{-4.5}}$& $2.71\times10^{-6}$& 
			$9.24\times10^{-13}$& 0.957& 0.0034&  $1.8\times{10^{-4}}$& 8.59\\
				11.64& 1.404&0.1&0.00891&0.958& ${10^{-4.5}}$& $2.71\times10^{-6}$& 
				$1.84\times10^{-13}$& 0.957& 0.0034&  $1.8\times{10^{-4}}$& 7.89\\
			\hline
			11.59& 1.40&1&0.0084&0.958& ${10^{-4.5}}$& $2.6\times10^{-6}$& 
			$1.64\times10^{-12}$& 0.956& 0.00299&  $1.9\times{10^{-4}}$& 8.84\\
			
			11.59& 1.40&0.5&0.0084&0.958& ${10^{-4.5}}$& $2.6\times10^{-6}$& 
			$8.2\times10^{-13}$& 0.956& 0.00299&  $1.9\times{10^{-4}}$& 8.54\\
				11.59& 1.40&0.1&0.0084&0.958& ${10^{-4.5}}$& $2.6\times10^{-6}$& 
				$1.64\times10^{-13}$& 0.956& 0.00299&  $1.9\times{10^{-4}}$& 7.84\\
			\hline
	\end{tabular}}
	
	\caption{ \small{Inflationary predictions of the model for various values of $\tilde{\lambda}$, $\xi$, $\delta$ and $\gamma$.  The number of e-folds is taken to be $N=60$.}}
	\label{mastertable}
\end{table}

\subsection{INFLATION ALONG S DIRECTION}

Here we briefly discussed  the case where the $S$ field has the r\^ole of the inflaton. In the potential (\ref{fullpotential}) we put $\langle\nu_H\rangle=0$ and 
$\langle \ov{\nu}_{H}\rangle=0$ so we have:
\begin{equation}
\begin{split}
V= \frac{144 \tilde{\kappa}^{2} S^{2}\left( \frac{m}{2 \tilde{\kappa}} - S^{2}\right)^{2}}{\left(3 - S^{2}\right)^{2}}.
\end{split}
\end{equation}
In order to remove the singularity of the denominator, we take $m=6 \tilde{\kappa}$. In this case we get the following simple form

\begin{equation}\label{S_chaotic}
V= 144 \tilde{\kappa} ^{2}S^{2}
\end{equation}

\noindent which is of the form of  a chaotic-potential.

Now the kinetic energy is defined as,
\begin{equation}
\begin{split}
\mathcal{L}=\frac{1}{2} K^{j}_{i} \left(\partial S\right)^{2}  -144 \tilde{\kappa} S^{2} \quad \text{where}
 \quad K^{j}_{i}=\frac{\partial^{2} K}{\partial S \partial S^{*}}=\frac{9}{\left(3-SS^{*}\right)^{2}}~\cdot 
\end{split}
\end{equation}
Let $S=\dfrac{X}{\sqrt{2}}$ then the potential in (\ref{S_chaotic}) becomes, $V= 72 \tilde{\kappa} ^{2}X^{2}$, and from the coefficient of the 
kinetic energy term we can find  $X$ in terms of a canonical normalized field $\chi$:
\begin{equation}
\begin{split}
 X =\sqrt{6} \tanh\left(\frac{\chi}{\sqrt{6}}\right).
\end{split}
\end{equation}
The potential in terms of the canonical normalized field reads as
\begin{equation}\label{ValongS}
V= 432 \tilde{\kappa}^{2}  \tanh^{2}\left(\frac{\chi}{\sqrt{6}}\right),
\end{equation}

\noindent which is analogous to the conformal chaotic inflation model or T-Model inflation already mentioned before. Potentials for the T-Model inflation are given in Equation (\ref{Tmodels}).  For $n=1$ the potential  become, $V(\chi)=\uplambda \tanh^{2}\left(\frac{\chi}{\sqrt{6}}\right)$, which is similar to our potential in (\ref{ValongS}) for $\uplambda=432 \tilde{\kappa}^{2}$. We can understand the inflationary behaviour in these type of models, by considering two cases. 

 First for $\chi\geqslant1$, by 
writing the potential in  exponential form we have

\begin{equation}
V= \uplambda \left(\frac{1-e^{-\sqrt{\frac{2}{3}} \chi}}{1+e^{-\sqrt{\frac{2}{3}} \chi}}\right)^{2}=\uplambda \left(1-\frac{2e^{-\sqrt{\frac{2}{3}} \chi}}{1+2e^{-\sqrt{\frac{2}{3}} \chi}}\right)^{2}=\uplambda\left(1-2e^{-\sqrt{\frac{2}{3}} \chi}\right)^{2}
\end{equation}

\noindent and for large values of $\chi$ we can write

\begin{equation}
V\simeq\uplambda\left(1-4e^{-\sqrt{\frac{2}{3}} \chi}\right)~,
\end{equation}

\noindent where $\uplambda=432 \tilde{\kappa}^{2}$. The slow roll parameters in terms of the field $\chi$ and for  large number of e-folds ($N$)  are

\begin{equation} \label{hN}
\frac{d\chi}{dN}=\frac{V^{\prime}}{V} =4\sqrt{\frac{2}{3}}e^{-\sqrt{\frac{2}{3}} \chi}.
\end{equation}

\noindent Integrating~(\ref{hN}) we have $\int{e^{\chi\sqrt{2/3}}d\chi}=\int{4\sqrt{\frac{2}{3}}dN}$, which gives the relation
\begin{equation}\label{nF}
\begin{split}
e^{-\sqrt{\frac{2}{3}}\chi} =\frac{3}{8 N}.
\end{split}
\end{equation}

\noindent Using the relation above  we have for the  slow-roll parameter $\epsilon$ that,
\begin{equation}
\begin{split}
\epsilon=\dfrac{1}{2}\left(\frac{V^{\prime}}{V}\right)^{2}=\dfrac{1}{2}\left(4\sqrt{\frac{2}{3}}e^{-\sqrt{\frac{2}{3}} \chi}\right)^{2}=\frac{3}{4 N^{2}}.
\end{split}
\end{equation}
Similarly the second slow-roll parameter $\eta$ is found to be,
\begin{equation}
\eta=\left(\frac{V^{\prime\prime}}{V}\right)=-\frac{1}{N}.
\end{equation}
Finally, the predictions for the tensor-to-scalar ratio $r$ and the natural-spectral index $n_{s}$ are,
\begin{equation}
\begin{split}
r=\frac{12}{N^{2}}\quad,\quad n_{s}=1+2\eta-6\epsilon=1-\frac{2}{N}-\frac{9}{4 N^{2}}
\end{split}
\end{equation}
\\
\noindent and for  $N=60$ e-foldings we get $n_{s} \simeq 0.9673$ and $r \simeq 0.0032$.

Regarding the case with $\chi \eqslantless 1$, we can see from the expression (\ref{ValongS}) that the potential reduces to a quadratic chaotic form. The tree-level inflationary predictions in this case are $\left(n_{s},r\right)\approx \left(0.967,0.130\right)$, which are ruled out with the latest \emph{Planck} $2015$ results. 

The discussion above strongly depends on the assumption $m=6\tilde{\kappa}$ that we imposed on the potential in order to simplify it. If we consider small variations of this assumption similar to \eqref{singularitycondition2} and modify the condition as, $m=6\tilde{\kappa}+\delta$, we will see that the parameter $\delta$ contributes only to $n_{S}$ while the tensor-to-scalar ratio $r$ remains constant.

\section{CONCLUSIONS}

In the present work we have studied ways to realise the inflationary scenario in a no-scale supersymmetric model 
 based on the Pati-Salam  gauge group $SU(4)\times SU(2)_L\times SU(2)_R$, supplemented with a $Z_2$ discrete  symmetry. The spontaneous  
 breaking of the group factor $SU(4)\to SU(3)\times U(1)_{B-L}$  is realised via the $SU(4)$ adjoint $\Sigma=(15,1,1)$  and the 
 breaking of the $SU(2)_{R}$ symmetry is achieved by non-zero vevs of the neutral components $\nu_{H}, \ov{\nu}_{H}$ of the  Higgs fields
  $(4,1,2)_H$ and $(\bar 4,1,2)_{\bar H}$. 

 We have considered a no-scale structure K\"ahler potential and  assumed that the  Inflaton field  is a combination of  
 $\nu_{H}, \ov{\nu}_{H}$  and find that the resulting potential is similar with the one presented in \cite{Ellis:2014dxa, Ellis:2016spb} 
 but our parameter space  differs substantially.  Consequently,  there are qualitatively  different solutions which are presented 
 and analysed in the present work. The results strongly depend on the parameter $\xi$ and for various characteristic values of the latter
 we obtain  different types of inflation models. In particular, for $\xi=0$ and canonical normalized field $\chi\geq{1}$, the potential 
 reduces to Starobinsky model and for $\xi=1$ the model receives a chaotic inflation profile. The results for $0<\xi<1$ have been analysed in detail while reheating via the decay of the inflaton in right-handed neutrinos is discussed.

 We also briefly discussed the alternative possibility where the $S$ field has the r\^ole of the inflaton. In this case, the potential is exponentially 
 flat for $\chi\geq{1}$.  Similar conclusions can be drawn for the Starobinsky model. On the other hand for small $\chi$ it reduces to  a quadratic potential.

In conclusion, the $SU(4)\times SU(2)_L\times SU(2)_R$ model described in this paper can provide inflationary predictions consistent with the observations. Performing a detailed analysis we have shown  that consistent solutions with the Planck data are found for a wide range of the parameter space of the model. In addition the inflaton can provide masses to the right-handed neutrinos and depending   on the value of reheating temperature and the right-handed
neutrino mass spectrum thermal
or non-thermal leptogenesis is a natural outcome. Finally we mention that, in several cases the tensor-to-scalar ratio $r$, a canonical measure of primordial gravity waves,  is close to$\sim{10^{-2}}-10^{-3}$ and can be tested in future experiments.

\vspace{1cm}
{\bf \large Acknowledgements}\quad\quad

\noindent The authors are thankful to George K. Leontaris, Qaisar Shafi, Tianjun Li and  Mansoor Ur Rehman for helpful discussions and useful comments. WA would like to thank the Physics Department at University of Ioannina for hospitality and for providing conducive atmosphere for research where part of this work has been carried out. WA  was  supported  by  the  CAS-TWAS  Presidents  Fellowship  Programme.

\end{document}